\documentclass[%
 reprint,
 superscriptaddress,
nofootinbib,
 amsmath,amssymb,
 aps,
 prx,
]{revtex4-2}

\usepackage{graphicx}
\usepackage{dcolumn}
\usepackage{bm}

\usepackage{xcolor}
\usepackage{amsmath}
\usepackage{comment}
\usepackage{mathrsfs}  

\usepackage{hyperref}

\usepackage{dsfont}
\usepackage{subfigure}

\newcommand{\ket}[1]{| #1 \rangle}
\newcommand{\bra}[1]{\langle #1 |}
\newcommand{\braket}[1]{\langle #1 \rangle}

\newcommand{\tr}{\text{tr}}

\begin{document}

\title{Quantum Hamiltonian-Based Models \& \\ the Variational Quantum Thermalizer Algorithm}

\author{Guillaume Verdon}
\thanks{Equal contribution}
\email[contact: ]{gverdon@x.team}
\affiliation{X, The Moonshot Factory, Mountain View, CA}
\affiliation{Institute for Quantum Computing, University of Waterloo, ON, Canada}
\affiliation{Department of Applied Mathematics, University of Waterloo, ON, Canada}

\author{Jacob Marks}
\thanks{Equal contribution}
\affiliation{X, The Moonshot Factory, Mountain View, CA}
\affiliation{Department of Physics, Stanford University, Stanford, CA}

\author{Sasha Nanda}
\affiliation{X, The Moonshot Factory, Mountain View, CA}
\affiliation{Division of Physics, Mathematics and Astronomy, California Institute of Technology, CA}

\author{Stefan Leichenauer}
\affiliation{X, The Moonshot Factory, Mountain View, CA}

\author{Jack Hidary}
\affiliation{X, The Moonshot Factory, Mountain View, CA}

\date{\today}

\begin{abstract}
We introduce a new class of generative  quantum-neural-network-based models called \textit{Quantum Hamiltonian-Based Models} (QHBMs). In doing so, we establish a paradigmatic approach for quantum-probabilistic hybrid variational learning, where we efficiently decompose the tasks of learning classical and quantum correlations in a way which maximizes the utility of both classical and quantum processors. In addition, we introduce the Variational Quantum Thermalizer (VQT) for generating the thermal state of a given Hamiltonian and target temperature, a task for which QHBMs are naturally well-suited. The VQT can be seen as a generalization of the Variational Quantum Eigensolver (VQE) to thermal states: we show that the VQT converges to the VQE in the zero temperature limit. We provide numerical results demonstrating the efficacy of these techniques in illustrative examples. We use QHBMs and the VQT on Heisenberg spin systems, we apply QHBMs to learn entanglement Hamiltonians and compression codes in simulated free Bosonic systems, and finally we use the VQT to prepare thermal Fermionic Gaussian states for quantum simulation.
\end{abstract}

\maketitle


\section{Introduction}

As near-term quantum devices move beyond the point of classical simulability, also known as quantum supremacy~\cite{boixo2018characterizing}, it is of utmost importance for us to discover new applications for Noisy Intermediate Scale Quantum devices~\cite{preskill2018quantum} which will be available and ready in the next few years. Among the most promising applications for near-term devices are Quantum Machine Learning (QML) \cite{farhi2018classification,mcclean2018barren,verdon2017quantum,dallaire2018quantum,mitarai2018quantum,biamonte2017quantum,verdon2019learning,khatri2019quantum,larose2019variational,schuld2018circuit,verdon2018universal,benedetti2019parameterized}, Quantum Simulation (QS) \cite{mcclean2016theory,peruzzo2014variational,kandala2017hardware,higgott2019variational,mcclean2019decoding,parrish2019quantum}, and Quantum-enhanced Optimization (QEO) \cite{farhi2014quantum,farhi2016quantum,jiang2017qaoa,verdon2019quantum, verdon2019learning,wang2019xy,brandao2018fixed,jiang2017near,li2019quantum}. Recent advances in these three areas have been dominated by a class of algorithms called hybrid quantum-classical variational algorithms. In these algorithms, a classical computer aids the quantum computer in a search over a parameterized class of circuits. These parameterized quantum circuits are sometimes called \textit{quantum neural networks} \cite{farhi2018classification,mcclean2018barren,verdon2017quantum}. Key to the success of quantum-classical algorithms is hybridization: in the near-term, quantum computers will be used as co-processors for classical devices. The work in this paper proposes a new way to hybridize certain quantum simulation and QML tasks in a way that fully takes advantage of the strengths of both devices.

The rise of variational quantum algorithms can be traced back to the invention and implementation of the Variational Quantum Eigensolver~\cite{peruzzo2014variational}, which sparked a Cambrian explosion of works in the field of near-term algorithms. Similarly, in this paper, not only do we introduce a direct generalization of the VQE, but we introduce the first member of a new class of algorithms, which we call \textit{quantum-probabilistic hybrid variational algorithms}, which are a combination of classical probabilistic variational inference~\cite{murphy2012machine,kingma2013auto,neal1993probabilistic,betancourt2017conceptual,rezende2015variational,louizos2017multiplicative,chen2018neural} and quantum variational algorithms.

More specifically, in this paper we focus on the task of generative modelling of \textit{mixed} quantum states. It is generally accepted that one must employ a quantum-based representation to efficiently learn the quantum correlations of a pure quantum state, as classical representations of quantum states scale poorly in the presence of quantum correlations such as entanglement~\cite{wilde2013quantum}. Mixed quantum states, arising from probabilistic mixtures of pure quantum states, generally exhibit both classical and quantum correlations. One must therefore learn a hybridized representation featuring both quantum correlations and classical correlations.

Within the new paradigm of quantum-probabilistic hybrid machine learning, we introduce a class of models called Quantum Hamiltonian-Based Models (QHBM). These models are constructed as thermal states (i.e., quantum exponential distributions) of a parameterized \textit{modular Hamiltonian}. As a first set of applications for this class of models, we explore the learning of unknown quantum mixed states, given access to several copies (quantum samples) of this quantum-probabilistic distribution.
As a second class of applications for QHBMs, we consider the task of generating the thermal state of a quantum system given knowledge of the system's Hamitonian.

Before proceeding with the main body of the paper, let us establish a broader context from the point of view of classical machine learning. The field of machine learning (ML) has seen several breakthroughs in recent years. These successes have often been attributed to the rapid advancements in \textit{deep learning} \cite{lecun2015deep}. In deep learning, one learns \textit{representations} of data distributions. Such representations can consist of a neural network~\cite{schmidhuber2015deep}, a tensor network~\cite{huggins2018towards,roberts2019tensornetwork}, or more generally a parameterized network of compositions of differentiable mappings. The parameters are \textit{trained} by optimizing some metric of success called the \textit{loss function}, often via gradient descent with backpropagation of errors~\cite{lecun1988theoretical}. One of the major tasks in \textit{unsupervised} deep learning is so-called generative modelling of a distribution.

In discriminative machine learning models, a model learns the \textit{conditional} probability of a target, $Y$, given the observation $x$, i.e., $P(Y|X = x)$. Generative models take on the task of learning the \textit{joint} probability distribution $P(X, Y)$, enabling a trained generative model to generate new data that \textit{looks} approximately like the training data. Generative models have been used for many unsupervised tasks, including generating new datapoints akin to a dataset, inpainting~\cite{yeh2017semantic}, denoising~\cite{vincent2008extracting}, superresolution~\cite{ledig2017photo}, compression~\cite{theis2017lossy}, and much more. Perhaps the most widespread generative techniques in classical machine learning are Generative Adversarial Networks (GANs)~\cite{goodfellow2014gan}, which pit a generative network and an adversarial network against each other, and variational autoencoders (VAEs) \cite{kingma2013auto}, which maximize a variational lower bound to the log likelihood of the data in order to increase the probability of the model to generate the dataset, and hence, similar datapoints. Both architectures have demonstrated great successes and have remained competitive with each other in terms of performance and scalability~\cite{razavi2019generating,brock2018large}.

Recently, a third type of generative algorithm --- the generalized form of an Energy Based Model (EBM) --- has been gaining traction in the classical machine learning community. This new architecture, derived as generalization of early energy-based architectures of neural networks such as Boltzmann machines, has been shown to be competitive with GANs and VAEs for generative tasks at large scales~\cite{du2019implicit}. The EBM approach draws its inspiration from distributions encountered in physics, namely thermal (exponential) distributions, where the probability of a given sample is proportional to the exponential of a certain function, called the \textit{energy}. Instead of sampling the exponential distribution from a fixed energy function, EBMs have a parameterization over a hypothesis space of energy functions, and the parameters which maximize the likelihood (relative entropy) to the dataset can be found via optimization. Boltzmann machines, a type of energy-based model with an energy function inspired by Ising models, have long been in use. The recent innovation, however, has been to use a neural network in order to have a more general and flexible parameterization over the space of energy functions. The differentiability of neural networks is then used to generate thermal samples according to Stochastic Langevin Gradient Dynamics~\cite{welling2011lang}. The construction presented in this paper is analogous to this energy-based construction, generalized to the quantum domain.

\section{Quantum Hamiltonian-Based Models}\label{sec:qhbm}

In order to represent the hybrid quantum-classical statistics of mixed states, the QHBM ansatz is structured in terms of a ``simple'' (i.e. nearly-classical) parameterized \textit{latent} mixed state which is passed through a unitary quantum neural network to produce a \textit{visible} mixed state. In this section we will introduce the general framework of QHBM before proceeding to training and examples in subsequent sections. The flow of classical and quantum information in a QHBM is illustrated in Figure~\ref{fig:qhbm_general}, and background on quantum neural networks is provided in Appendix~\ref{app:qnn}.

The variational \textit{latent} distribution $\hat{\rho}_{\bm{\theta}}$ with \textit{variational parameters} $\bm{\theta}$ is constructed in such a way that the preparation of $\hat{\rho}_{\bm{\theta}}$ is simple, from a quantum computational standpoint.\footnote{We define this more precisely a few paragraphs below when giving examples of structures for the latent space distribution.} Quantum correlations are incorporated through the unitary $ \hat{U}(\bm{\phi})$ with \textit{model parameters} ${\bm{\phi}}$.

Our complete variational mixed state is then
\begin{align}
\hat{\rho}_{\bm{\theta\phi}} = \hat{U}(\bm{\phi}) \hat{\rho}_{\bm{\theta}}\hat{U}^\dagger(\bm{\phi}). 
\end{align}
We call this state the variational \textit{visible state}. It is the output of the inference mechanism of our composite model, and will be either a thermal state or learned approximation to a target mixed state, depending on the task.

\subsection{Modular Hamiltonians and the Exponential Ansatz}

Quantum Hamiltonian-Based Models are quantum analogues of classical energy-based models, an analogy which will be made clear in this section.

Without loss of generality, we can consider the latent distribution to be a thermal state of a certain Hamiltonian: 
 \begin{align}\label{eq:latent_distr_ansatz}
 \hat{\rho}_{\bm{\theta}} = \tfrac{1}{\mathcal{Z}_{\bm{\theta}}}e^{-\hat{K}_{\bm{\theta}}},\quad \mathcal{Z}_{\bm{\theta}} = \text{tr}[e^{-\hat{K}_{\bm{\theta}}}].
 \end{align}
We call this $\hat{K}_{\bm{_\theta}}$ the \textit{latent modular Hamiltonian}, and it is one of a class of operators parameterized by the latent variational parameters $\bm{\theta}$.
Here $\mathcal{Z}_{\bm{\theta}} = \text{tr}[e^{-K_{\bm{\theta}}}]$ is the \textit{model partition function}, which is, notably, also parameterized by the latent variational parameters. Now, given this form for the latent state, notice that the variational visible state is a thermal state of a related Hamiltonian: 
\begin{equation}\label{eq:visible_ansatz}
\hat{\rho}_{\bm{\theta\phi}} =  \tfrac{1}{\mathcal{Z}_{\bm{\theta}}}e^{-\hat{U}(\bm{\phi})\hat{K}_{\bm{\theta}}\hat{U}^\dagger(\bm{\phi})} \equiv \tfrac{1}{\mathcal{Z}_{\bm{\theta}}}e^{-\hat{K}_{\bm{\theta \phi}}} ,
\end{equation}
where we define the \textit{model modular Hamiltonian} as $\hat{K}_{\bm{\theta \phi}}\equiv \hat{U}(\bm{\phi})\hat{K}_{\bm{\theta}}\hat{U}^\dagger(\bm{\phi})$, which is parameterized by both the latent variational parameters $\bm{\theta}$ and the model parameters $\bm{\phi}$. Thus, we see that our QHBM ansatz represents a class of quantum distributions which are thermal states of parameterized Hamiltonians. As we will see below, this exponential structure is useful for computing relative entropies of the model against target distributions.

We note that the above structure is in direct analogy with classical energy-based models. In such models, the variational distribution is of the exponential form
\(p_{\bm{\theta}}(\bm{x}) = \tfrac{1}{\mathcal{Z}_{\bm{\theta}}}e^{-E_{\bm{\theta}}(\bm{x})},\) where \( \mathcal{Z}_{\bm{\theta}}\equiv \sum_{\bm{x}} e^{-E_{\bm{\theta}}(\bm{x})}\)
and the energy function $E_{\bm{\theta}}(\bm{x})$ is parameterized by a neural network. The network is trained so that the samples from $p_{\bm{\theta}}$ mimic those of a target data distribution. 

In place of a parameterized classical energy function, we have a parameterized modular Hamiltonian operator, and the variational model is a thermal state of this operator. This justifies why we call it a Quantum Hamiltonian-Based Model instead of simply an Energy-Based model. The thermal state of the Hamiltonian is designed to replicate the quantum statistics of the target data.

We will distinguish two distinct tasks within the QHBM framework. The first is generative learning of a target mixed state, given access to copies of said mixed state. We call this task \textit{Quantum Modular Hamiltonian Learning} (QMHL). As described above, we should think of this task as finding an effective modular Hamiltonian for the target state. The second task involves being given a target Hamiltonian, and generating approximate thermal states of some temperature with respect to that Hamiltonian. In our framework, this means that we variationally learn an effective modular Hamiltonian which reproduces the statistics of the target thermal state. We call this task \textit{Variational Quantum Thermalization} (VQT). We will treat QMHL and VQT in depth in Sections~\ref{sec:QMHL} and \ref{sec:VQT}, respectively.

Before discussing in more detail the structure of the latent space, let us draw attention to two quantities which will be important for the training process in both VQT and QMHL: the partition function and the Von Neumann entropy of the model. Comparing equations \eqref{eq:latent_distr_ansatz} and \eqref{eq:visible_ansatz}, we see that the partition function $\mathcal{Z}_{\bm{\theta}}$ is the same for both the latent state and the visible state. Furthermore the Von Neumann entropy of the latent and visible states are also identical due to the invariance of Von Neumann entropy under unitary action, 
\begin{align}\label{eq:entropy_cons}
S(\hat{\rho}_{\bm{\theta}}) = S\left(\hat{U}(\bm{\phi}) \hat{\rho}_{\bm{\theta}}\hat{U}^\dagger(\bm{\phi})\right) = S( \hat{\rho}_{\bm{\theta}\bm{\phi}}).
\end{align}
We will leverage this in our training of the QHBM, where efficient computation or estimation of the latent state entropy will give us knowledge of the visible state entropy.

\begin{figure}[ht]
\centering
\includegraphics[width=0.45\textwidth]{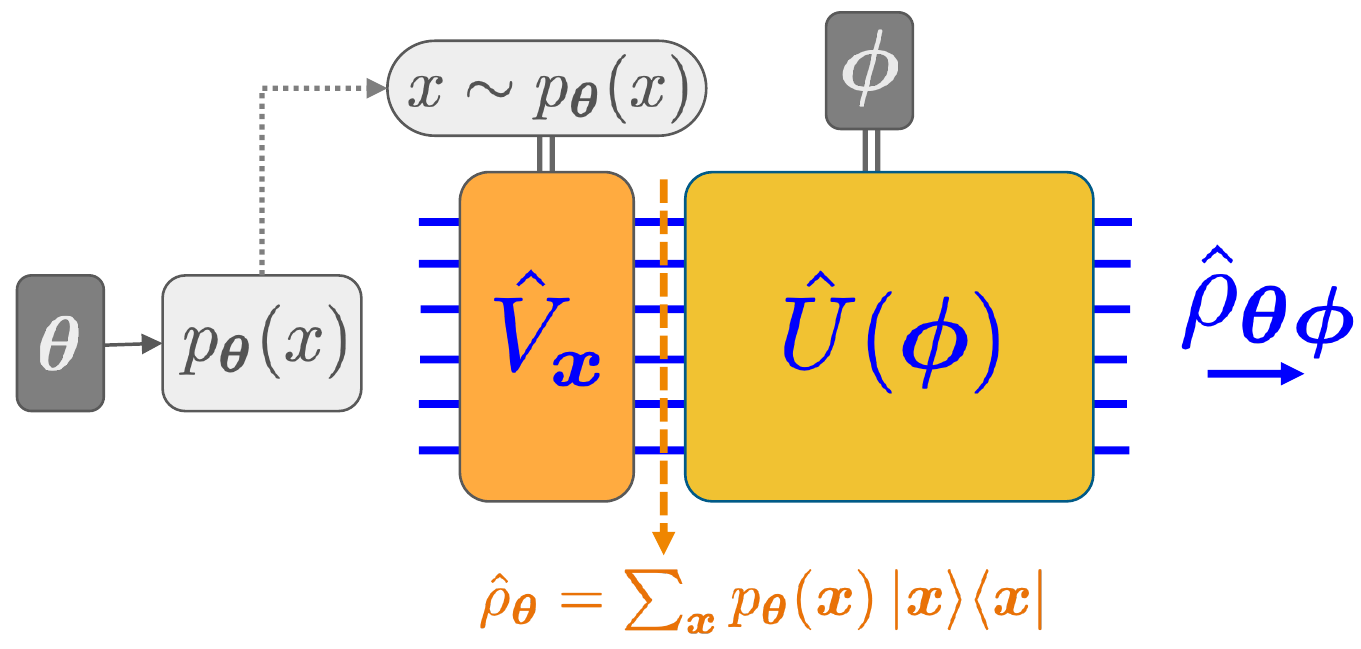}\caption{Quantum-classical information flow diagram for hybrid quantum-probabilistic inference for a Quantum Hamiltonian-Based Model with a general classical latent distribution. Here, we have unitaries $\hat{V}_{\bm{x}}$ which map the quantum computer's initial state $\ket{\bm{0}}$ to computational the computation basis state $\ket{\bm{x}}$ which corresponds to the sampled value $\bm{x} \sim p_{\bm{\theta}}(\bm{x})$ on a given run. }
\label{fig:qhbm_general}
\end{figure}

\subsection{Structure of the Latent Space}

In this section we will discuss the form of the latent model $\hat{\rho}_{\bm{\theta}}$. We consider a good choice of a latent space ansatz to be one that is \textit{quantumly simple}, i.e., of low-complexity for the quantum computer to prepare. The two types of latent distributions employed in this paper will be either a \textit{factorized latent state}, or a \textit{general classical latent space distribution}. Let us now introduce these two cases and dive further into the specifics of each.

\subsubsection{Factorized Latent Space Models}

A first choice for the latent space structure is a \textit{factorized latent state}, which is a latent state of the the form
\begin{align}
\hat{\rho}_{\bm{\theta}} = \bigotimes_{j=1}^N\hat{\rho}_j(\bm{\theta}_j),
\end{align}
where the total quantum system is separated into $N$ smaller-dimensional subsystems, each with their own set of variational parameters $\bm{\theta}_j$ for which the mixed states of the subsystems $\hat{\rho}_j(\bm{\theta}_j)$ are uncorrelated. 

This structure has several useful benefits. Notice that, due to this tensor product structure, the latent modular Hamiltonian becomes a sum of modular Hamiltonians of the subsystems,
\begin{align}\label{eq:factorized}
\hat{K}_{\bm{\theta}} \equiv \sum_j \hat{K}_j(\theta_j),\quad \hat{\rho}_{\bm{\theta}} = \bigotimes_{j=1}^N\tfrac{1}{\mathcal{Z}_{\bm{\theta}_j}}e^{-\hat{K}_j(\theta_j)}, \end{align}
where $\mathcal{Z}_{\bm{\theta}_j} = \text{tr}[e^{-\hat{K}_j(\theta_j)}]$ is the $j^{\text{th}}$ subsystem partition function, and $\hat{K}_j(\theta_j)$ the $j^{\text{th}}$ subsystem modular Hamiltonian. This sum decomposition becomes useful when estimating expectation values of the modular Hamiltonian, as it becomes a sum of expectation values of the subsytems' modular Hamiltonians,
\begin{align}
\braket{\hat{K}_{\bm{\theta}}}= \sum_j \braket{\hat{K}_j(\theta_j)}.\end{align}
From the above expression, we see that the partition function is a product of the subsystem partition functions $\mathcal{Z}_{\bm{\theta}}  = \prod_j \mathcal{Z}_{\bm{\theta}_j} $, and hence the logarithm of the partition function becomes a sum:
\begin{align}\log(\mathcal{Z}_{\bm{\theta}}) = \textstyle\sum_{j=1}^N \log(\mathcal{Z}_{\bm{\theta}_j}).\end{align}
Furthermore, the entropy of the latent state (and hence of the visible state, via equation \eqref{eq:entropy_cons}) becomes additive over the entropies of the subsystems:
\begin{align}S(\hat{\rho}_{\bm{\theta}}) = \textstyle \sum_{j=1}^N S\Big(\hat{\rho}_j(\bm{\theta}_j)\Big)\end{align}
This is convenient, as estimating $N$ entropies of states in $d$-dimensional Hilbert space is much simpler generally than computing the entropy of a state in $d^N$-dimensional space.

Another feature of a factorized state is that the number of parameters used to describe such a distribution is linear in the number of subsystems $N$. The precise number of parameters depends on the structure of the states within each subsystem. There are many possibilities, and we will present some concrete examples below.

Finally, by learning a completely decorrelated (in terms of both entanglement and classical correlations) representation in latent space, we are effectively learning a representation which has a natural orthogonal basis for its latent space, allowing for latent space interpolation, independent component analysis, compression code learning, principal component analysis, and many other spin-off applications. In classical machine learning, this is known as a \textit{disentangled} representation~\cite{bengio2013representation}, and there have been several recent works adressing this machine learning task~\cite{burgess2018understanding}.

\subsubsection{Examples of Latent Subsystem Modular Hamiltonians}
For the QHBMs considered in this paper, as explored in-depth in Section~\ref{sec:apps_n_exps}, we use several different types of latent subsystem modular Hamiltonians within the factorized-latent-state framework of equation~\eqref{eq:factorized}. We will review them here.

The first class of modular Hamiltonians employed in our experiments are qudit operators \cite{Marks2017qudit} diagonal in the computational basis. In our particular numerics in Section~\ref{sec:heis}, we use two-level systems, i.e. qubits. The more general qudit case is akin to the eigenbasis representation of the Hamiltonian of a multi-level atom, \begin{align}\hat{K}_j(\bm{\theta}_j)  = \textstyle\sum_{k=1}^{d_j} \theta_{jk} \ket{k}\!\bra{k}_j,\end{align}
where the the eigenvalues $\{\theta_{jk} \}_{k=1}^{d_j}$ of this Hamiltonian form the set of variational parameters of the distribution. The latent distribution of each subsystem is then effectively a softmax~\cite{koller1999general} function of the eigenvalues of this Hamiltonian, and can be considered as equivalent to a general $N$-outcome multinoulli~\cite{murphy2012machine} distribution. The total number of parameters for such a latent space parameterization with factorized subsystems scales as the sum of the subsystem dimensions, $\sum_{j=1}^N d_j$, which is much smaller than the most general (non factorized) latent distribution, whose number of parameters scales as the product $\prod_{j=1}^N d_j$, which is the total dimension of the system's Hilbert space.

The second type of modular Hamiltonian we use, in Section~\ref{sec:compression}, is the number operator of a continuous-variable quantum mode (qumode), or harmonic oscillator, \begin{align}\hat{K}_j(\theta_j)= \tfrac{\theta_j}{2}\, (\hat{x}^2 + \hat{p}^2)= \theta_j \, \left(\hat{a}_j^\dagger \hat{a}_j +\tfrac{1}{2}\right).\end{align}
The exponential distribution of such modular Hamiltonians then becomes a single-mode thermal state~\cite{serafini2017quantum} which has a Wigner phase space representation as a symmetric Gaussian. The single parameter per mode here, $\theta_j$, modulates the variance of the Gaussian. This single-mode thermal state is the closest thing we have to a latent product of Gaussians, which is the standard choice of latent distributions in several variational inference models in classical machine learning, including variational autoencoders~\cite{kingma2013auto}. Gaussian states of such quantum modes are very natural to prepare on continuous-variable quantum computers~\cite{lloyd1999quantum}, and are also emulatable on digital quantum computers~\cite{verdon2018universal}.

Finally, as will be explored in Section~\ref{sec:fermion}, our third type of latent subsystem modular Hamiltonian is the particle number operator for Fermions, $\hat{K}_j(\theta_j) = \theta_j\hat{c}_j^\dagger \hat{c}_j$.


\subsubsection{General Classical Latent Space Models}

In some cases, a decorrelated (disentangled) representation for the latent space is not possible or capable of yielding accurate results. In classical deep learning, this factorized latent space prior assumption has been one of the main reasons that VAEs with uncorrelated Gaussian priors perform worse than GANs on several tasks. To remedy this situation, several classical algorithms for more accurate variational inference of latent space distributions have popped up, notably including Neural ODEs~\cite{chen2018neural}, normalizing flow-based models~\cite{rezende2015variational}, and neural energy-based models~\cite{du2019implicit}.

This same generalization can be applied to QHBMs, moving from the factorized latent space structure of \eqref{eq:factorized} to more general distributions. This should be understood as a delegation of parts of the representation to a classical device running classical probabilistic inference. The job of the classical device is to sample from the classical latent space distribution.

In the most general formulation, the latent state can be represented as 
\begin{align}\hat{\rho}(\bm{\theta}) = \sum_{x\in\Omega} p_{\bm{\theta}}(\bm{x})\ket{\bm{x}}\!\bra{\bm{x}},
\end{align} where the summation is formal, and runs over the index set $\Omega$ of some basis in Hilbert space, which we call the computational basis. Here the probability distribution $p_{\bm{\theta}}(\bm{x})$ is generally a parameterized class of variational probability distributions over the computational basis domain $\Omega$. Since a general categorical (multinoulli) distribution would need a number of parameters which scales as $|\Omega|$, the dimension of the global Hilbert space, which is generally exponentially large (e.g., $|\Omega| = 2^N$ for $N$ qubits), one needs a more efficient parameterization of the latent distribution. This is where classical algorithms for probabilistic and variational inference come into play.

The classical computer is thus tasked with the variational learning of this classical distribution. There are a plethora of techniques for variational inference to choose from, as were listed above. The key feature we need from such classical algorithms is the ability to estimate the log-likelihood of the model distribution, the gradients of the (log) partition function, and the entropy. These quantities will become useful in various generative learning tasks, see Section~\ref{sec:training} for more details.

The most natural fit for our needs are the modern variant of classical energy-based models \cite{du2019implicit}. In EBMs, a neural network $E_{\bm{\theta}}:\Omega \mapsto \mathbb{R}$ parameterizes the energy map from the computational basis to a real value, i.e. for any value $\bm{x}\in\Omega$ it can produce the corresponding energy $E_{\bm{\theta}}(\bm{x})$. Furthermore, due to the easy differentiability of neural networks, one can rather easily compute gradients of the energy function, $\nabla_{\bm{\theta}} E_{\bm{\theta}}(\bm{x})$ and $\nabla_{\bm{x}} E_{\bm{\theta}}(\bm{x})$. To leverage these gradients, one uses Langevin dynamics or Hamiltonian Monte Carlo to generate samples $x\sim p_{\bm{\theta}}(x) $ according to the exponential distribution of the energy function,
\begin{align}p_{\bm{\theta}}(x) = \tfrac{1}{\mathcal{Z}_{\bm{\theta}}} e^{-E_{\bm{\theta}}(\bm{x})}\end{align}
where $\mathcal{Z}_{\bm{\theta}} \equiv \sum_{\bm{x}\in\Omega}e^{-E_{\bm{\theta}}(\bm{x})} $ is the partition function. As a straightforward feature of this approach, given an ensemble of points sampled from a different distribution $\bm{x}\sim q(\bm{x})$, one can straightforwardly use the neural network to evaluate expectation values of the energy with respect to this distribution, \(\mathbb{E}_{\bm{x}\sim q(\bm{x})}[E(\bm{x})] \). Additionnally, this approach can provide an estimate and/or gradients of the entropy of the distribution $S(p_{\bm{\theta}}) = \sum_{\bm{x}\in \Omega} -p_{\bm{\theta}}(\bm{x}) \log p_{\bm{\theta}}(x)$. Given the abilities of the classical model above, one can use the classical algorithm to generate the latent mixed state $\hat{\rho}(\bm{\theta}) =\sum_{x\in\Omega} p_{\bm{\theta}}(\bm{x})\ket{\bm{x}}\!\bra{\bm{x}}$ by sampling $\bm{x}\sim p_{\bm{\theta}}$ and preparing the computational basis state $\ket{\bm{x}}\!\bra{\bm{x}}$ after each sample. Furthermore, we can define the modular Hamiltonian for this latent variational model as
\begin{align}\hat{K}_{\bm{\theta}} = \sum_{\bm{x}\in\Omega} E_{\theta}(\bm{x})\ket{\bm{x}}\!\bra{\bm{x}}.\end{align}
Thus one can estimate expectation values of this operator by feeding standard basis measurement results through the neural network function $E_{\bm{\theta}}(\bm{x})$ and averaging. We now have outlined all the ingredients needed for inference and training of this type of hybrid QHBM.

Generally, given a target mixed state $\hat{\sigma}_D$ which we want to generatively model, we know that there exists a diagonal representation of this mixed state:
\begin{equation}\forall\,\hat{\sigma}_{\mathcal{D}},\  \exists\, \hat{W}_{\mathcal{D}}\, :\ \hat{W}^\dagger_{\mathcal{D}}\hat{D}_{\mathcal{D}}\hat{W}_{\mathcal{D}} = \hat{\sigma}_{\mathcal{D}}, \, \hat{D}_{\mathcal{D}} = \textstyle\sum_{\bm{x}}\lambda_{\bm{x}}\ket{\bm{x}}\!\bra{\bm{x}} \end{equation} where $\tr(\hat{D}_{\mathcal{D}}) = 1$, and $W_{\mathcal{D}}$ is a unitary operator. Thus the approach outlined above has, in principle, the capacity to represent any mixed state. The challenge remains to pick a proper prior for both the unitary quantum neural network $\hat{U}(\bm{\phi})$ and for the parameterization of the classical latent distribution $p_{\bm{\theta}}(\bm{x})$ (or equivalently, the latent energy function $E_{\bm{\theta}}(\bm{x})$). A good ansatz is one which uses knowledge about the physics of the system to form a prior for both the latent space and the unitary transformation. Although it may be tempting to use a general multinoulli distribution for the latent distribution combined with a universal random quantum neural network for the unitary ansatz, the model capacity in this case is far too large. One encounters not only exponential overhead for parameter estimation of the high-dimensional multinoulli distribution, but also the quantum version of the no-free lunch theorem~\cite{mcclean2018barren}. In Section~\ref{sec:apps_n_exps}, we see several scenarios where the structure of the latent space and the unitary transformation are well-adapted to the complexity of the situation at hand.

\section{Quantum Generative Modelling of Mixed \& Thermal States}\label{sec:training}

In this section, we introduce two types of generative tasks for quantum Hamiltonian-based models which are also valid for more general quantum-probabilistic models. The first task is quantum modular Hamiltonian learning (QMHL), where one effectively learns the logarithm (modular Hamiltonian) of an unknown data mixed state. Once trained, one is then able to reproduce copies of this unknown mixed state via an exponential (thermal) distribution of the learned modular Hamiltonian. The second tasks is dual to the first task: given a known Hamiltonian, learn to produce copies of the thermal state of this 
Hamiltonian. While both tasks can be considered in the broad class of generative modelling, the first task is a quantum machine learning task, while the second may be considered as a quantum simulation task.

As we have already detailed how to construct several variants of the QHBM with a variety of latent space structure in Section \ref{sec:qhbm} --- including details on how to compute expectations of the modular Hamiltonian, generate samples of the latent state, and compute quantities such as the partition function and entropy of the model --- we will focus on the general framework of the QHML and VQT without too many specifics about how to compute the various quantities involved. For more details on gradients of the loss functions introduced here, see Appendix \ref{app:qhbm_grad}.

\subsection{Quantum Modular Hamiltonian Learning of Mixed Quantum States}\label{sec:QMHL}
Analogous with the classical case, quantum generative models seek to learn to replicate the statistics and correlation structure of a given quantum data distribution from which we have a finite number of samples. QHBMs make this learning process feasible by positing a form that inherently separates the tasks of learning the quantum correlations in the learned representation of a quantum distribution. 

Typical classical distributions are a set $\mathcal{D}$ of samples drawn from some underlying distribution $d\sim p_{\mathcal{D}}$. The sampled dataset distribution is simply the categorical distribution over the datapoints $d\in\mathcal{D}$. A quantum dataset generally can be a similar mix of various datapoints, in this case quantum states $\hat{\sigma}_d$ observed at different frequencies (inverse probabilities). The effective mixed state representing this dataset is then \(
 \hat{\sigma}_{\mathcal{D}} = \sum_{d\in \mathcal{D}} p_d\,\hat{\sigma}_d,\) where we obtain the mixed state $\hat{\sigma}_d$ with probability $p_d$ from our dataset $\mathcal{D}$. As modelling a mixture of states is effectively equivalent to the task of modelling one mixed state, we will focus on simply learning to approximate a single \textit{data mixed state} $\hat{\sigma}_{\mathcal{D}}$ using a variational quantum-probabilistic model. We assume that we have access to several copies of this data mixed state directly as quantum data available in quantum memory.

\begin{figure}[ht]
\centering
\includegraphics[width=0.34\textwidth]{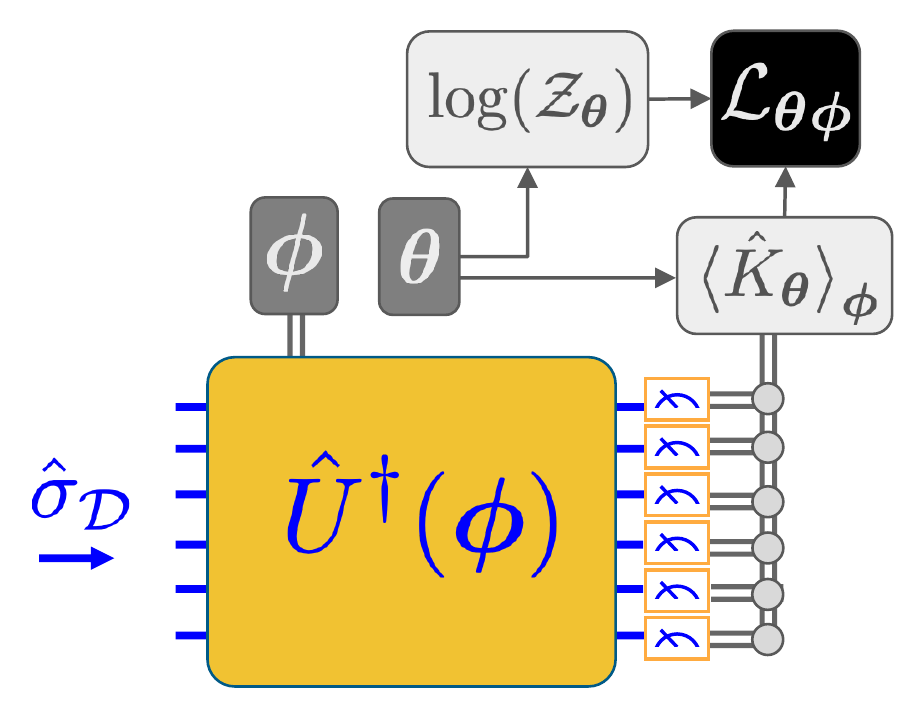}\caption{Information flow for the process of training for the QHBM applied to Modular Hamiltonian Learning. Grayscale indicates classical information processing happening on a classical device while colored registers and operations are stored and executed on the quantum device. The $\bm{\theta}$ parameters determine the latent space distribution and thus the modular Hamiltonian $\hat{K}_{\bm{\theta}}$. From the known latent distribution, one can estimates of the parameterized partition function $\log\mathcal{Z}_{\bm{\theta}}$ on the classical device. One then applies the inverse unitary quantum neural network $\hat{U}^\dagger(\bm{\phi})$ and estimates the expectation value of the modular Hamiltonian $\hat{K}_{\bm{\theta}}$ at the ouput via several runs of inference on the quantum device and measurement. The partition function and modular expectation are then combined to yield the quantum variational cross entropy loss function \eqref{eq:qmhl_loss}.}
\label{fig:modham_flow}
\end{figure}

Thus, in complete generality, we consider the problem of learning a mixed state $\hat{\sigma}_{\mathcal{D}}$, and let $\hat{\rho}_{\bm{\theta\phi}}$ be our model's candidate approximation to $\hat{\sigma}_{\mathcal{D}}$, where $\bm{\theta,\phi}$ are variational parameters. In classical probabilistic machine learning, one minimizes the Kullback-Leibler divergence (relative entropy), an approach known as ``expectation propagation.'' \cite{murphy2012machine}  We directly generalize this method to quantum probabilistic machine learning, now aiming to minimize the quantum relative entropy between the quantum data distribution and our quantum model's candidate distribution, subject to variations of the parameters. Thus we aim to find
\begin{align}
\text{argmin}_{\bm{\theta,\phi}}D(\hat{\sigma}_{\mathcal{D}}\Vert\hat{\rho}_{\bm{\theta\phi}}),
\end{align}
where $D(\hat{\sigma}\Vert\hat{\rho}) = \text{tr}(\hat{\sigma} \log \hat{\sigma}) - \text{tr}(\hat{\sigma} \log \hat{\rho} )$ is the quantum relative entropy. Due to the positivity of relative entropy, the optimum of the above cost function is achieved if and only if the variational distribution of our model is equal to the quantum data distribution:
\begin{align}\label{eq:rel_ent_var}D(\hat{\sigma}_{\mathcal{D}}\Vert\hat{\rho}_{\bm{\theta\phi}})\geq 0, \quad D(\hat{\sigma}_{\mathcal{D}}\Vert\hat{\rho}_{\bm{\theta\phi}})= 0 \iff \hat{\sigma}_{\mathcal{D}}=\hat{\rho}_{\bm{\theta\phi}}.\end{align}
We can use this as a variational principle: the relative entropy is our loss function and we can find optimal parameters of the model $\{\bm{\theta}^*,\bm{\varphi}^*\}$ such that \(\hat{\rho}_{\bm{\theta}^*\bm{\phi}^*}\approx \hat{\sigma}_{\mathcal{D}}\). As we will see below, the use of QHBMs becomes crucial for the tractability of evaluation of the relative entropy, making its use as a loss function possible. Details on the gradient computation of the loss are discussed in Appendix~\ref{app:qhbm_grad}.

Now, for a target quantum data mixed state $\hat{\sigma}_{\mathcal{D}}$ and a QHBM variational state $\hat{\rho}_{\bm{\theta\phi}}$ of the form described in Equation \eqref{eq:visible_ansatz}, our goal is to minimize the forward relative entropy,
\begin{equation}D(\hat{\sigma}_{\mathcal{D}} \Vert \hat{\rho}_{\bm{\theta\phi}}) = -S(\hat{\sigma}_{\mathcal{D}})+  \mathrm{tr}(\hat{\sigma}_{\mathcal{D}}\hat{K}_{\bm{\theta}\bm{\phi}}) + \log(\mathcal{Z}_{\bm{\theta}}). \end{equation}
 Notice that the first term, $S(\hat{\sigma}_{\mathcal{D}}) = -\mathrm{tr}(\hat{\sigma}_{\mathcal{D}}\log \hat{\sigma}_{\mathcal{D}})$, the entropy of the data distribution, is independent of our variational parameters, and hence, for optimization purposes, irrelevant to include. This is convenient as the entropy of the dataset is a priori unknown. We can thus use the last two terms as our loss function for QHML. These terms are known as the \textit{quantum cross entropy} between our data state and our model, 
 \begin{align}
\mathcal{L}_{\textsc{qmhl}}(\bm{\theta,\phi}) \equiv - \text{tr}(\hat{\sigma}_{\mathcal{D}} \log \hat{\rho}_{\bm{\theta}\bm{\phi}} ) = \mathrm{tr}(\hat{\sigma}_{\mathcal{D}}\hat{K}_{\bm{\theta}\bm{\phi}}) + \log(\mathcal{Z}_{\bm{\theta}}).
\end{align}
We call this loss function the \textit{quantum variational cross entropy loss}. The first term can be understood as the expectation value of the model's modular Hamiltonian with respect to the data mixed state. In order to estimate this modular energy expectation, it is first useful to express the modular Hamiltonian of the model in terms of the latent modular Hamiltonian: $\hat{K}_{\bm{\theta,\phi}} = \hat{U}(\bm{\phi})\hat{K}_{\bm{\theta}}\hat{U}^\dagger(\bm{\phi})$. Then, by the cyclicity of the trace, we can rewrite the loss as \begin{align}\label{eq:qmhl_loss}
\mathcal{L}_{\textsc{qmhl}}(\bm{\theta,\phi}) = \mathrm{tr}([\hat{U}^\dagger(\bm{\phi})\hat{\sigma}_{\mathcal{D}}\hat{U}(\bm{\phi})]\hat{K}_{\bm{\theta}}) + \log(\mathcal{Z}_{\bm{\theta}}).
\end{align}
To better understand this, we define the \textit{pulled-back data state} $\hat{\sigma}_{\mathcal{D},\bm{\phi}} \equiv \hat{U}^\dagger(\bm{\phi})\hat{\sigma}_{\mathcal{D}}\hat{U}(\bm{\phi})$, which is the state obtained by feeding our quantum data through the unitary quantum neural network circuit in reverse (or rather, inverse). Thus, this term in the loss is equivalent to the expectation value of the latent modular Hamiltonian with respect to the pulled-back data, $\braket{\hat{K}_{\bm{\theta}}}_{\hat{\sigma}_{\mathcal{D},\bm{\phi}}}$. The flow of quantum and classical information for the task of Modular Hamiltonian Learning is depicted in Figure~\ref{fig:modham_flow}. See Section~\ref{sec:qhbm} for details on estimating the partition function, depending on the chosen latent space structure.

Finally, an interesting feature to note about about this choice of cross-entropy loss function is that in the limit where the model approximates the data state, i.e,. when the relative entropy in \eqref{eq:rel_ent_var} converges to zero, then the loss function itself converges to the entropy of the data:
\begin{equation}
\mathcal{L}_{\textsc{qmhl}}(\bm{\theta,\phi}) \overset{\hat{\rho}_{\bm{\theta\phi}}\rightarrow \hat{\sigma}_{\mathcal{D}}}{\longrightarrow} S(\hat{\sigma}_{\mathcal{D}})
\end{equation}
This means that, as a by-product of the learning process, after convergence of the training of our model, we are automatically provided with an estimate of the entropy of our data distribution simply by observing what the loss value has converged to. This has wide-ranging implications about the potential use of QHBM and QMHL, combined with quantum simulation, to allow us to estimate entropies and various information theoretic quantities using quantum computers. More on this in the discussion in Section \ref{sec:disc_fw}.

\subsection{Variational Quantum Thermalizer Algorithm: Quantum Simulation of Thermal States}\label{sec:VQT}

In this section we formally introduce the Variational Quantum Thermalizer (VQT) algorithm, the free energy variational principle behind it, how it relates to the Variational Quantum Eigensolver (VQE) \cite{peruzzo2014variational} in a certain limit, and how one can use QHBMs for this task.

The variational quantum thermalization task can be described as the following: given a Hamiltonian $\hat{H}$ and a target inverse temperature $\beta = 1/T$, we wish to generate an approximation to thermal state
\begin{align}
\hat{\sigma}_{\beta} = \tfrac{1}{\mathcal{Z}_{\beta}}e^{-\beta \hat{H}},\quad  \mathcal{Z}_\beta = \text{tr}(e^{-\beta \hat{H}}).
\end{align}

Here $\mathcal{Z}_\beta$ is known as the thermal partition function. Our strategy will be to phrase this quantum simulation task as a quantum-probabilistic variational learning task. Suppose we have a quantum-probabilistic ansatz for this thermal state, \(\hat{\rho}_{\bm{\theta\phi}}\) with parameters $\{\bm{\theta},\bm{\phi}\}$. We can consider the relative entropy between this unknown thermal state and our variational model,
\begin{align}
    D(\hat{\rho}_{\bm{\theta\phi}}\Vert \hat{\sigma}_\beta) &= -S(\hat{\rho}_{\bm{\theta\phi}}) - \text{tr}(\hat{\rho}_{\bm{\theta\phi}}\log \hat{\sigma}_\beta)\\ \nonumber &=-S(\hat{\rho}_{\bm{\theta\phi}}) +\beta \text{tr}(\hat{\rho}_{\bm{\theta\phi}}\hat{H}) +\log \mathcal{Z}_\beta.
\end{align}
This is known as the \textit{quantum relative free energy} of our model with respect to the target Hamiltonian $\hat{H}$. The reason it is called a \textit{relative} free energy is because it is the difference of free energy $F(\hat{\xi}) \equiv \tr(\hat{H}\hat{\xi}) - \tfrac{1}{\beta}S(\hat{\xi})$, up to a factor of $\beta$, between our ansatz state and the true thermal state,
\[D(\hat{\rho}_{\bm{\theta\phi}}\Vert \hat{\sigma}_\beta)  = \beta F(\hat{\rho}_{\bm{\theta\phi}}) - \beta F(\hat{\sigma}_{\beta}) .\]
Further note that the positivity of relative entropy implies that the minimum of free energy is achieved by the thermal state;
\[D(\hat{\rho}_{\bm{\theta\phi}}\Vert \hat{\sigma}_\beta)  = 0 \implies F(\hat{\rho}_{\bm{\theta\phi}}) = F(\hat{\sigma}_{\beta}) \text{ and } \hat{\rho}_{\bm{\theta\phi}}=\hat{\sigma}_{\beta}.\]
Thus, given a variational ansatz for the thermal state, by minimizing the free energy as our loss function, 
\[\mathcal{L}_{\textsc{vqt}}(\bm{\theta},\bm{\phi}) =  \beta F(\hat{\rho}_{\bm{\theta\phi}}) =\beta \text{tr}(\hat{\rho}_{\bm{\theta\phi}}\hat{H}) -S(\hat{\rho}_{\bm{\theta\phi}}),  \]
we find optimal parameters $\{\bm{\theta}^*,\bm{\varphi}^*\}$ such that \(\hat{\rho}_{\bm{\theta}^*\bm{\phi}^*}\approx \hat{\sigma}_{\mathcal{\beta}}\).

\begin{figure}[ht]
\centering
\includegraphics[width=0.45\textwidth]{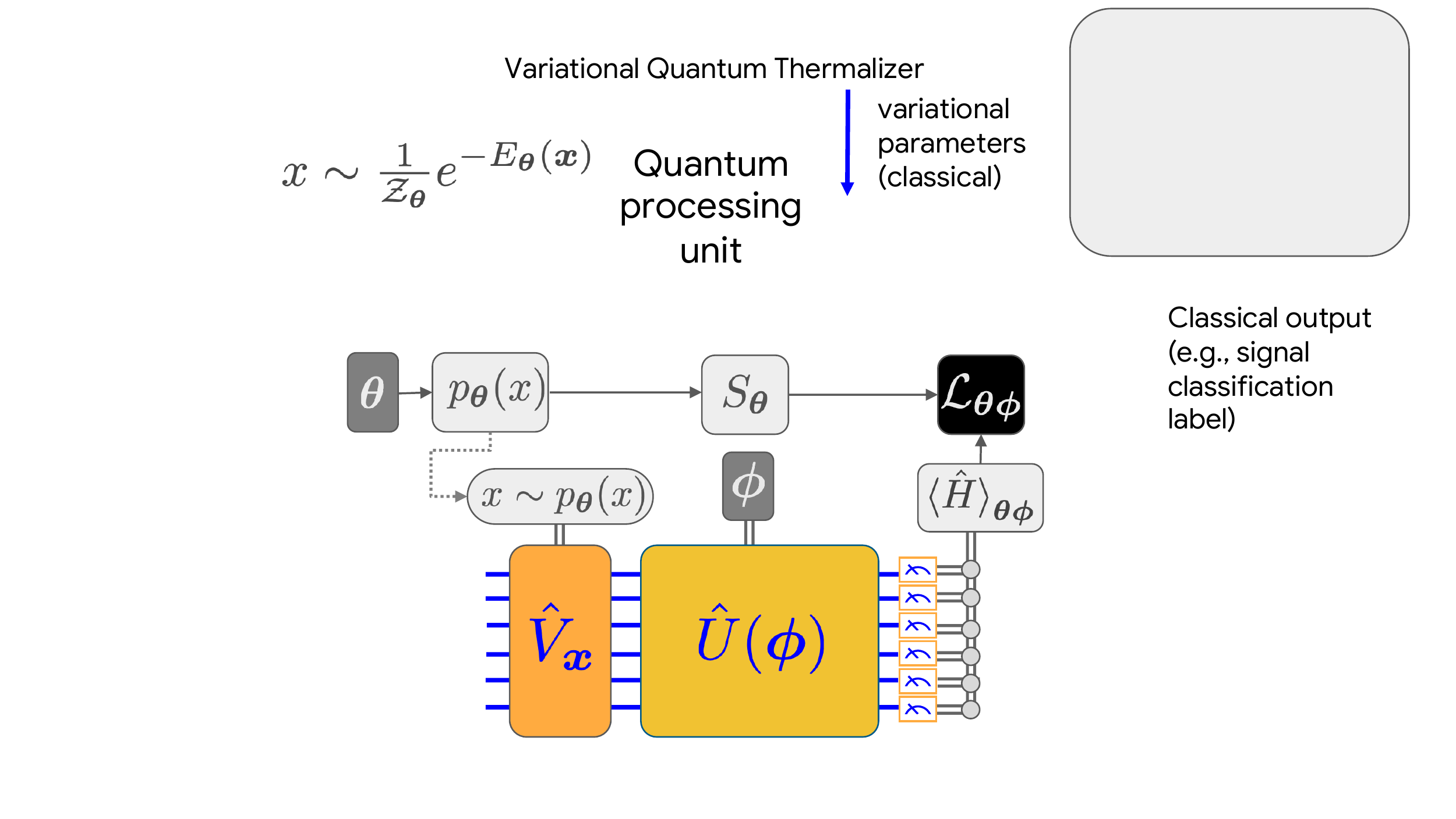}\caption{Information flow for the process of inference and training for the QHBM applied to VQT. Grayscale indicates classical information processing happening on a classical device while colored registers and operations are stored and executed on the quantum device. Here we focus on a general case for the latent variational distribution. The $\bm{\theta}$ parameters determine the latent space distribution. From this distribution, one can compute the entropy $S_{\bm{\theta}}$ classically. Using samples from the latent distribution $x\sim p_{\bm{\theta}}(x)$, one applies a quantum operation to prepare the state $\ket{x}\!\bra{x}$ via the unitary $\hat{V}_x$ from the initial state of the quantum device. One then applies the unitary quantum neural network $\hat{U}(\bm{\phi})$ and estimates the expectation value of the Hamiltonian $\hat{H}$ at the output via several runs of classical sampling of $x$ and measurement. The entropy and energy expectation are then combined into the free energy loss for optimization.}
\label{fig:vqt_flow}
\end{figure}


Now that we have our loss function, we can briefly examine how one could use a QHBM ansatz for the above. The great advantage of the QHBM structure here is that the entropy of the variational model distribution is that of the latent distribution, as pointed out in equation \eqref{eq:entropy_cons}. As the entropy of the latent variational distribution is stored on the classical computer and assumed to be known a priori, only the evaluation of the energy expectation (the first term in the above) requires the quantum computer. Thus, the number of runs required to estimate the loss function should be similar to the number of runs required for the Variational Quantum Eigensolver and other variational algorithms whose loss only depends on expectation values \cite{mcclean2016theory}. 

Note that, in the case where our state converges to the true thermal state upon convergence of the training, then the value of our loss function will give us the free energy of the thermal state, which is proportional to the log of the thermal partition function:
\[\mathcal{L}_{\textsc{vqt}}(\bm{\theta},\bm{\phi}) \overset{\hat{\rho}_{\bm{\theta\phi}}\rightarrow \hat{\sigma}_{\beta}}{\longrightarrow} \beta F(\hat{\sigma}_{\beta}) = -\log \mathcal{Z}_\beta.\]
This is effectively a variational free energy principle, and the basis of the VQT.

As a final note, let us see how we recover the Variational Quantum Eigensolver and its variational principle in the limit of low temperature. If we divide our loss function by $\beta$, i.e., directly minimizing the free energy, 
$\tilde{\mathcal{L}}(\bm{\theta},\bm{\phi})  = \tfrac{1}{\beta}\mathcal{L}_{\textsc{vqt}}(\bm{\theta},\bm{\phi})  = F(\hat{\rho}_{\bm{\theta\phi}})$, then in the limit of zero temperature,
\[\tilde{\mathcal{L}}(\bm{\theta},\bm{\phi}) \overset{\beta\rightarrow \infty}{\longrightarrow} \braket{\hat{H}}_{\bm{\theta\phi}},  \]
we recover the loss function for the ground state Variational Quantum Eigensolver (VQE), and thus we recover the ground state variational principle. Note that in the limit of zero temperature, there is no need for latent space parameters as there is no entropy and the latent state is unitarily equivalent to any starter pure state. 

Thus, the VQT is truly the most natural generalization of the VQE to non-zero temperatures states.

\section{Applications \& Experiments}\label{sec:apps_n_exps}

\begin{table*}[]
\caption{QHBM Experiments Ingredients}
\label{tab:equiv}
\begin{tabular}{l@{\hskip 0.25in}l@{\hskip 0.25in}l@{\hskip 0.25in}l}
\hline
\textbf{Physical System} & \textbf{Qubits (Spins)} & \textbf{Gaussian Bosons} & \textbf{Gaussian Fermions} \\ 
\hline
\textbf{Base Quantities} \rule{0pt}{3ex} & $\hat{S}_i, \hat{S}_j$ & $\hat{x}_j, \hat{p}_k$ & $\hat{c}_{2 j - 1},  \hat{c}_{2 j}$ \\ 
\hline
\textbf{Representation} \rule{0pt}{3ex}  & $\hat{\rho}$ & $\Gamma_{B} = \frac{1}{2}Tr(\hat{\rho} \{\hat{\bm{\xi}},  \hat{\bm{\xi}}^T\})$ & $\Gamma_{F} = \frac{i}{2}Tr(\hat{\rho} [\hat{\bm{\xi}},  \hat{\bm{\xi}}^T])$ \\ \hline
\textbf{Ansatz} \rule{0pt}{3ex} & $\hat{\rho}_{\bm{\theta} \bm{\phi}} = \hat{U}^{\dag}(\bm{\phi}) \hat{\rho}(\bm{\theta}) \hat{U}(\bm{\phi})$ & $S^{T}(\bm{\phi}) \Gamma_{\bm{\theta}}S(\bm{\phi}) $ & $O^{T}(\bm{\phi}) \Gamma_{\bm{\theta}}O(\bm{\phi})$ \\ 
\hline
\textbf{Latent Ansatz} \rule{0pt}{4ex} & $\hat{\rho}(\bm{\theta}) = \displaystyle\bigotimes_{j = 1}^{N} \begin{pmatrix} 1 - p_j(\theta_j) & 0 \\ 0 & p_j(\theta_j) \end{pmatrix}$ & $\Gamma_{\bm{\theta}} = \displaystyle\bigoplus_{j = 1}^{N} \begin{pmatrix}\nu_j(\theta_j) & 0 \\0 & \nu_j(\theta_j) \end{pmatrix}$ & $\Gamma_{\bm{\theta}} = \displaystyle\bigoplus_{j = 1}^{N} \begin{pmatrix} 0 & \lambda_j(\theta_j) \\ -\lambda_j(\theta_j) & 0 \end{pmatrix}$ \\ 
\hline

\end{tabular}
\end{table*}

Our framework is general enough to apply in many situations of interest in quantum computing~\cite{nielsen2002quantum}, quantum communication~\cite{gisin2007quantum}, quantum sensing~\cite{degen2017quantum}, and quantum simulation~\cite{Kassal_2008, Kassal_2011, Whitfield_2011, B804804E, Lanyon_2010, Aspuru-Guzik1704, Lamm_2018, Endres200, Cheneau_2012, PhysRevLett.115.035302, PhysRevLett.89.156801, PhysRevLett.106.107402, Latta2011QuantumQO, Kaufman_2016, Berman_1991chaos, Sieberer_2019}. In this section we focus on applications to quantum simulation and quantum communication, which illustrate the QHBM frameworks for spins/qubits, Bosonic, and Fermionic systems.

In the Bosonic and Fermionic examples, we restrict ourselves to Gaussian states for two purposes: first, these highlight the ways in which our framework is well-suited to utilizing problem structure; and second, classical methods exist for simulating quantum systems restricted to these types of states, so they serve as a verification and benchmarking of our methods. This allows us to reach system sizes which would not be simulable on classical computers were we to simulate the wavefunction in the Hilbert space directly. However, we emphasize that Quantum Hamiltonian-Based Models apply to a much broader class of quantum states.

Gaussian states are the class of thermal (and ground) states of all Hamiltonians that are quadratic in creation and annihiliation operators (in second quantization). On the Fermionic side, this class contains a variety of tight-binding models, including topological models like the Su-Schrieffer-Heeger (SSH) model~\cite{ssh1, ssh2, ssh3}, and mean-field BCS superconducting states. Bosonic Gaussian states include coherent and squeezed states, and already enable (when coupled with measurement) a plethora of applications in quantum communication, including quantum teleportation and quantum key distribution. In addition, Gaussian states are interesting in their own right, and play a significant role in quantum information theory~\cite{bravyi2004lagrangian, bravyi2011classical, Melo_2013}, as well as quantum field theory in curved space-time \cite{dewitt1975quantum} and quantum cosmology~\cite{mukhanov2007introduction}. 

Gaussian states are special in that our latent-space factorization postulate is justified physically in addition to holding numerically. As a consequence of Williamson's theorem~\cite{williamson1936}, we are guaranteed that for any thermal state, there exists a Gaussian transformation that decouples all subsystems in latent space (a procedure often referred to as normal mode decomposition~\cite{qcvbook}). For Gaussian systems, we can also work directly with $2 N \times 2 N$ covariance matrices rather than density matrices.

\subsection{Variational Quantum Thermalizer for the 2D Heisenberg Model}
\label{sec:heis}
As a demonstration of the effectiveness of VQT for spin systems (qubits), we learn a thermal state of the two-dimensional Heisenberg model, which exhibits spin-frustration despite its simplicity. To be precise, we consider the model 
\begin{align}
    \hat{H}_{\textsc{heis}} = &\sum_{\langle ij\rangle_h} J_{h} \hat{\bm{S}}_i \cdot \hat{\bm{S}}_j +  \sum_{\langle ij\rangle_v} J_{v} \hat{\bm{S}}_i \cdot \hat{\bm{S}}_j
\end{align}
where $h$ ($v$) denote horizontal (vertical) bonds, and $\langle \cdot \rangle $ represent nearest-neighbor pairs. As laid out in Section~\ref{sec:qhbm}, we use a factorized latent space distribution. Even though we are considering a two-dimensional model, we simulate VQT using an imagined one-dimensional array of qubits. As such, we parameterize our quantum neural network with only single qubit rotations and two-qubit rotations between adjacent qubits in our simulated one-dimensional quantum computer.

In particular, our paramterized unitary is composed out of three layers of gates. Each layer consists of an arbitrary single qubit rotation of the form $\exp[{i(\phi^{1}_j\hat{X}_j + \phi^2_j\hat{Y}_j + \phi^3_{j}\hat{Z}_j )}]$ applied to each qubit, and a two-qubit rotation  $\exp[{i(\phi^{4}_j\hat{X}_j\hat{X}_{j+1} + \phi^5_j\hat{Y}_j\hat{Y}_{j+1} + \phi^6_{j}\hat{Z}_j\hat{Z}_{j+1} )}]$ applied to a complete set of neighboring qubits. The neighbors acted upon by two-qubit gates are staggered in sequential layers. No parameters are shared within or between layers. So in the first layer, the first and second qubits, third and fourth, and so on, are coupled, while in the second layer it is the second and third qubits, fourth and fifth, etc., which are coupled. In addition, we do not assume periodic boundary conditions for our qubits, and as such do not include any gates between the first and last qubits.

While we use relative entropy as our cost function, we also look at the trace distance and state fidelity of the reconstructed target states, as defined in Appendix~\ref{app:dist}. As the modular Hamiltonian is not unique, in modular Hamiltonian learning we choose to compare the thermal states generated from target and learned Hamiltonian. In Appendix~\ref{app:ansatz}, we investigate the performance of the factorized latent space ansatz on systems that in general do not factorize.

\begin{figure}[ht]
\centering
\includegraphics[width=0.45\textwidth]{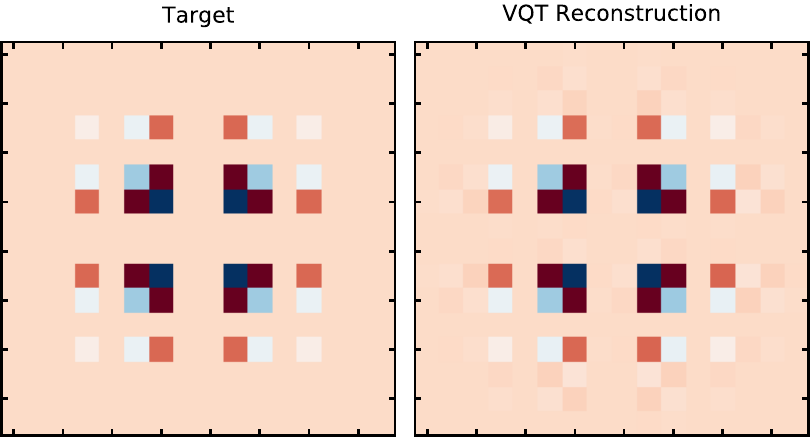}
\caption{Density Matrix visualizations for thermal state of a two-dimensional Heisenberg model. Left: Target thermal state of Hamiltonian; Right: VQT reconstruction after 200 training steps. Specific model parameters are $N_x = N_y = 2$, at $\beta = 2.6, \quad J_x = 1.0, \quad J_y = 0.6$. Performance is sustained for larger systems, but harder to visualize due to sparsity of density matrix elements.}
\label{fig:VQT_dm}
\end{figure}

\begin{figure}[ht]
\centering
\includegraphics[width=0.45\textwidth]{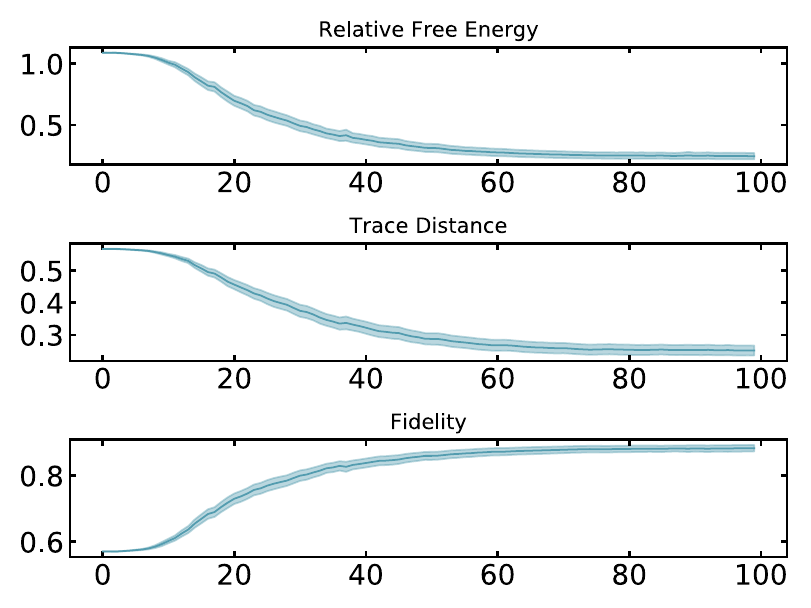}
\caption{Performance metrics for VQT ansatz reconstruction of thermal state of two-dimensional anti-ferromagnetic Heisenberg model with $N_x = N_y = 2$, at $\beta = 0.5, \quad J_x = 1.0, \quad J_y = 1.0$. Upper: Training loss (Free energy of ansatz minus free energy of target state; Center: trace distance; Lower: mixed state fidelity. Solid line denotes mean, and shaded region represents $95\%$ confidence interval. Model trained on $100$ randomly chosen sets of initial parameters picked uniformly from the unit interval.}
\label{fig:VQT_CI}
\end{figure}

In Fig.~\ref{fig:VQT_dm}, we show the density matrix elements of the target thermal state, and the reconstructed density matrix elements learned from the corresponding Hamiltonian via VQT after 200 variational steps. A few iterations later, the differences become too small to visualize and then the procedure converges.

Figure~\ref{fig:VQT_CI} shows mean values and $95\%$ confidence intervals for three performance metrics over $100$ VQT training iterations, for $100$ randomly initialized sets of variational parameters. This illustrates that our model trains generically and is not strongly dependent on initial parameters. Furthermore, while our model was only trained to minimize relative entropy, it is worth noting that it successfully learns with respect to trace distance and fidelity.

\subsection{Learning Quantum Compression Codes}
\label{sec:compression}
Continuous Variable (CV) systems offer an intriguing alternative to discrete variable quantum information processing~\cite{serafini2017quantum}. In this context, our parameterized ansatz finds a decoupled representation of subsystem of quantum modes in latent space, and the entropy is effectively described by the modes' effective temperature. High temperature modes contribute most to the entropy, and low temperature modes contain relatively little information. This decorrelated structure could be used to devise quantum approximate compression codes.

We recall that for Bosonic Gaussian states, the problem of dimensionality is made tractable by working with the phase space quadratures $\hat{x}$ and $\hat{p}$ defined in Appendix~\ref{app:cv}. The real symmetric covariance matrix $\Gamma_{B}$ associated to a mixed state $\hat{\rho}$ is given by the matrix elements
\begin{align}
\Gamma^{ab}_{B} = \frac{1}{2}\mathrm{Tr} (\hat{\rho} \{\hat{\xi}^a, \hat{\xi}^b\}),
\end{align}
where each component of $\hat{\bm{\xi}}$ is one of the $\hat{x}$ or $\hat{p}$ quadratures.

We also remind the reader that Gaussian transformations act as symplectic transformations on the covariance matrix, and that the symplectic transformation that diagonalizes $\Gamma_{B}$, 

\begin{align}
    S \Gamma_{B} S^T = \bigoplus_{j = 1}^{N_b}\begin{pmatrix}
 \nu_j & 0 \\
0 & \nu_j
\end{pmatrix},
\end{align}
also diagonalizes the Hamiltonian. In this context, modular Hamiltonian learning consists of finding the symplectic diagonalization of the Hamiltonian given quantum access to the covariance matrix. For more details on Bosonic Gaussian Quantum Information, see Appendix~\ref{app:cv}.

To keep our method as close to implementation as possible, we directly parameterize the space of symplectic matrices in terms of the effective action of single mode phase shifts and squeezers, and two-mode beam splitters as detailed in Appendix~\ref{app:param}, and in~\cite{killoran2018continuousvariable}.

We consider a translationally invariant linear harmonic chain
\begin{align}
    \hat{H}_{\textsc{lhc}} = \sum_j \omega \hat{x}_j^2 + \hat{p}_j^2 + 2 \chi \hat{x}_j \hat{x}_{j+1},
\end{align}
where $\hat{x}_{N+1} = \hat{x}_{1}$, as in ~\cite{qcvbook}. At the point $\omega = 2 |\chi|$, the system is critical and the operator $\hat{H}_{\textsc{lhc}}$ becomes unbounded from below.

First, we find the ground state of a long chain (we use 200 sites) close to criticality. We then cut the chain at two points and keep one of the two parts of the bipartition, performing a partial trace over the rest of the system. Given this reduced density matrix for the subsystem, we use QMHL to find an approximate representation of a modular Hamiltonian that would generate this as a thermal state. 

\begin{figure}[ht]
\centering
\includegraphics[width=0.48\textwidth]{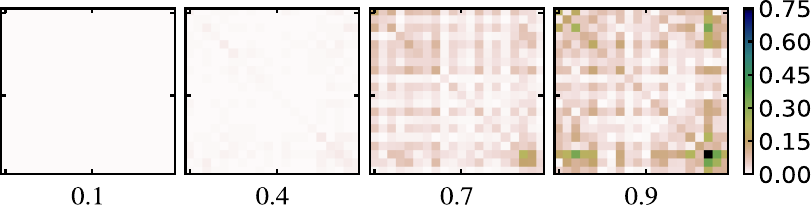}
\caption{Compression of latent space representation for reduced state of harmonic chain near criticality ($\omega = 1.0, \, \chi = 0.499$), with Modular Hamiltonian Learning. Subsystem of 10 modes traced out from full chain of length 200. Plots visualize absolute value of difference between reconstructed and true covariance matrix elements, at single-shot compression ratios of 0.1, 0.4, 0.7 and 0.9. Color scale is relative to largest matrix element of the true covariance matrix.}
\end{figure}

Our ansatz for QHBM is perfectly suited for latent space compression. The entropy of our ansatz is only dependent on the latent space parameters, where we have uncoupled oscillators at different effective temperatures. Compression consists of systematically setting low-temperature oscillators to their ground state.

\begin{figure}[htb]
\centering
\includegraphics[width=0.48\textwidth]{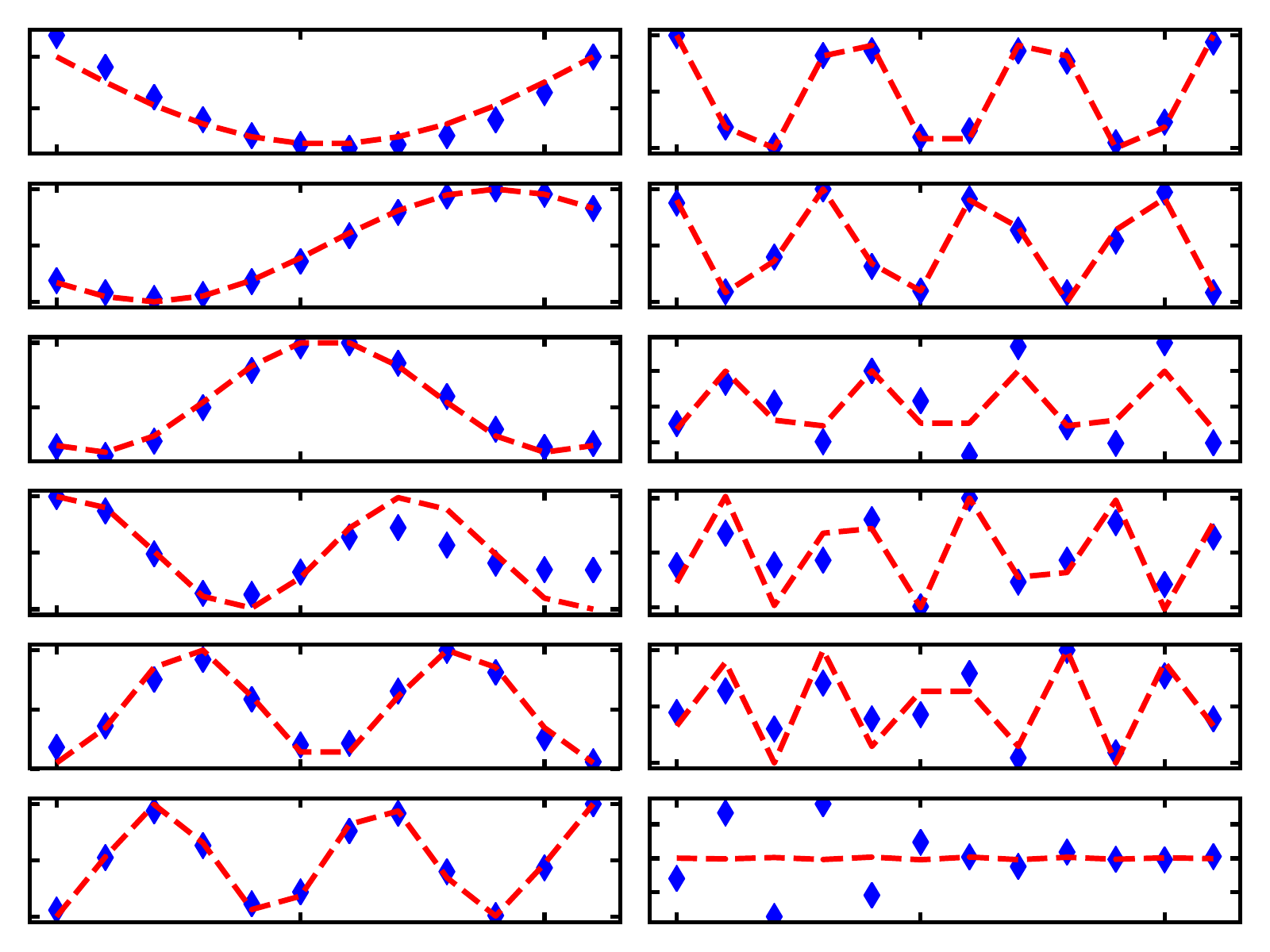}
\caption{Modular modes of reduced harmonic chain. Blue diamonds denote Modular Hamiltonian Learning results, and dashed red line denotes exact result from Williamson decomposition. Modes arranged in order of increasing eigenvalue (from top to bottom, and from left to right). System under consideration was a harmonic chain of length 100 lattice sites (subsystem of 12 sites), with $\omega = 1$, $\chi = -0.2$.}
\label{fig:mod_modes}
\end{figure}

By running our circuit forward (encoding), compressing, and then backward (decoding), we see that for the reduced state of the chain, even when we remove almost all of the latent mode information, we can still approximately reconstruct the original state.

In the case of Bosonic Gaussian systems, the columns of the diagonalizing symplectic transformation are the modular modes of the reduced system. Thus, QHBM are essentially directly learning these modes. In Fig.~\ref{fig:mod_modes}, we show modular mode reconstructions (after shifting and rescaling) for a  translationally invariant harmonic chain away from criticality.

\subsection{Preparation of Fermionic Thermal States}
\label{sec:fermion}

In recent years, great progress has been achieved in using quantum systems to simulate quantum dynamics. This quantum simulation has been achieved in a variety of physical implementations, including ultracold atomic gases in optical lattices \cite{bloch2005ultracold}, trapped ions \cite{kielpinski2002architecture}, and superconducting circuits \cite{barends2014superconducting}. QHBM are applicable to any of these implementations, where VQT could be used to prepare an approximate thermal state. Moreover, modular Hamiltonian learning could facilitate the simulation of out-of-equilibrium time-evolution, going beyond what is feasible with classical methods. 

As a proof of principle of the application of QHBM to these problems, we use the VQT to learn an efficient preparation of d-wave superconducting thermal states. The BCS mean-field theory Hamiltonian is quadratic in creation and annihilation operators, so we restrict our attention to Fermionic Gaussian states. These states are useful in a variety of quantum chemistry applications, where they are often desired as initial states for further quantum processing routines \cite{openfermion}. They have also been used to model impurities in metals~\cite{Bravyi2017, PhysRevX.6.031045}, which are key to understanding much of the collective phenomena in strongly-correlated materials~\cite{PhysRev.124.41, PhysRevA.97.022707, PhysRevLett.120.083401}. 

In analogy with the Bosonic Gaussian case considered above, we remind the reader that for Fermionic Gaussian systems, the real anti-symmetric covariance matrix $\Gamma_{F}$ associated to a mixed state $\hat{\rho}$ is given by the matrix elements
\begin{align}
\Gamma^{ab}_{B} = \frac{i}{2}\mathrm{Tr} (\hat{\rho} [\hat{\xi}^a, \hat{\xi}^b]),
\end{align}
where now $\hat{\bm{\xi}} = (\hat{c}_1, \ldots, \hat{c}_{2N})$ is a vector of Majorana operators derived from the system's physical creation and annihilation operators.

Fermionic Gaussian transformations act as orthogonal transformations on covariance matrices, and the same orthogonal that diagonalizes the covariance matrix,
\begin{align}
    O \Gamma_{F} O^T = \bigoplus_{j = 1}^{N_f} \begin{pmatrix}
 0 & - \lambda_j \\
\lambda_j & 0
\end{pmatrix},
\end{align}
also diagonalizes the Hamiltonian. For more details, see Appendix~\ref{app:gaussfermion}.

The VQT uses minimization of relative entropy to learn the orthogonal transformation concurrently with the effective frequencies of the decoupled Fermionic modes, essentially learning a representation of the Bogoliubov transformation. However, implementing this on a discrete-variable quantum computer requires the additional initialization steps of applying a qubit-to-Fermion transformation, such as the Jordan-Wigner transformation, and then transforming to the basis of Majorana modes. The orthogonal transformation itself can be directly parameterized in terms of Givens rotations~\cite{givens} on pairs of modes. This in turn allows for the analytic computation and propagation of variable gradients through the quantum circuit.

\begin{figure}[t]
\centering
\includegraphics[width=0.48\textwidth]{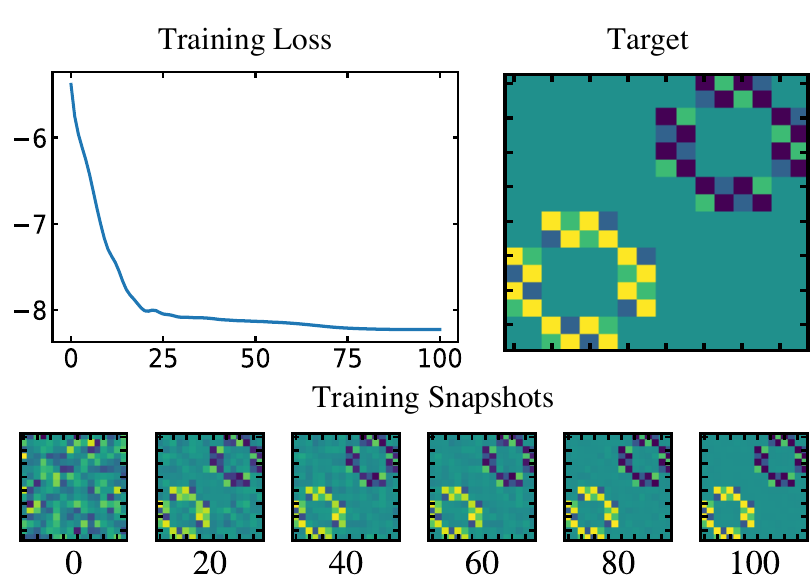}
\caption{VQT training for \textit{d}-wave thermal state with $N_x = N_y = 2$, with $\beta = 1,\, \quad t = 0.3, \, \quad \Delta = 0.2$. Upper left: VQT training loss (free energy) verus training iteration. Upper right: Depiction of matrix elements of target covariance matrix in Majorana basis. Below: Snapshot visualization of matrix elements of variationally parameterized covariance matrix in Majorana basis at various points in training process.}
\label{fig:dwave}
\end{figure}

For the completely general BCS Hamiltonian and interpretations of its constituent terms, we refer the reader to Appendix~\ref{app:dwave}. For present purposes, we consider the case of uniform Cooper pairing, i.e. $\Delta_{ij} = \Delta$, and without loss of generality we set the chemical potential to zero, $\mu = 0$, leading to a Hamiltonian of the form

\begin{align}
   \hat{H}_{d_{x^2-y^2}} = & -t \sum_{\langle i,j\rangle, \sigma}( \hat{a}^{\dag}_{i, \sigma} \hat{a}_{j, \sigma}  + \hat{a}^{\dag}_{j, \sigma} \hat{a}_{i, \sigma})    \\
   & \Delta \sum_{\langle i,j\rangle}  (\hat{a}^{\dag}_{i, \uparrow} \hat{a}^{\dag}_{j, \downarrow} - \hat{a}^{\dag}_{i, \downarrow} \hat{a}^{\dag}_{j, \uparrow} + \mathrm{h.c.}) \nonumber.
\end{align}

While the \textit{d}-wave Hamiltonian is Gaussian, the Cooper pairing induces terms such as  $\hat{a}^{\dag}_{1, \uparrow} \hat{a}^{\dag}_{2, \downarrow}$. As a result, we switch to the Majorana basis before learning a state preparation circuit. See Appendix~\ref{app:gaussfermion} for details. In Figure~\ref{fig:dwave}, we illustrate the training of VQT on a typical d-wave thermal state. The snapshots on the bottom are covariance matrices output by the partially trained QNN at intervals of 20 learning steps. After 150 steps, the target covariance matrix and VQT reconstruction are indiscernible by fidelity, entropy and trace distance. We note that in classical simulations of QHBM for Gaussian Fermions, our models were able to learn successfully for much larger systems (up to 50 Fermions), but the covariance matrices became too large to clearly discern individual matrix elements as in Figure~\ref{fig:dwave}.

\section{Discussion \& Future Work}\label{sec:disc_fw}

Recent literature in quantum machine learning and quantum-classical variational algorithms has focused on the learning of pure quantum state approximations to the ground or excited states of quantum systems. If we wish to simulate and model the physics of realistic quantum systems, we must allow for the possibility of interplay between classical and quantum correlations. 

In this paper we introduced the general formalism of QHBMs. Previous techniques for learning mixed quantum states have been tailored specifically for low-rank density matrices \cite{larose2019variational}. Our method, which minimizes relative entropy rather than Hilbert-Schmidt distance, is generic and can be applied to mixed and thermal states of any rank. It has the additional benefits of yielding estimates of mixed state entropy, free energy, and the diagonalizing transformation of the target system, the last of which enables modular time evolution and facilitates full quantum simulation of a previously unknown system. This further opens up the possibility of using quantum machine learning to compute state entropies of analytically intractable systems. Considering the vast body of literature exploring the computation of entropies in physical quantum systems \cite{casini2009entanglement}, this novel application of quantum-classical hybrid computing could be quite significant.

Latent space factorization, while not obligatory for QHBM, leads to widespread applications based on learning decorrelated (factorized) quantum representations, akin to classical \textit{disentangled representations}~\cite{bengio2013representation}. In the context of unsupervised learning, learning a disentangled latent space representation could be useful for tasks such as latent space interpolation~\cite{bojanowski2017optimizing} and latent space vector arithmetic, quantum principal component analysis~\cite{wold1987principal}, clustering~\cite{hofmann2001unsupervised}, and more generally entropy-based quantum compression as demonstrated in Section~\ref{sec:compression}. Such compression could also be used for unsupervised pre-training in discriminative learning \cite{erhan2010does}. Progressively learning to compress layer-wise while growing the depth of the unitary quantum neural network would yield a procedure akin to \textit{layer-wise pretraining} in classical machine learning~\cite{erhan2010does}. Compression could also be used for denoising quantum data, something previous approaches to quantum autoencoders were not suitable for \cite{romero2017quantum}.

As for the case of using a diagonal latent space Hamiltonian for QMHL, we are then effectively learning the eigenbasis of the data density matrix. 

The great advantage over previously proposed variational quantum algorithms for quantum state diagonalization is that our method employs the relative entropy rather than a Hilbert-Schmidt metric, thus allowing us to learn to diagonalize much higher rank quantum states. Furthermore, learning to diagonalize a quantum density matrix is essentially related to the Quantum Principal Component Analysis algorithm \cite{lloyd2014quantum} for classical data, and other related quantum machine learning algorithms \cite{biamonte2017quantum,rebentrost2014quantum}. Our approach could be seen as a variational alternative method for these algorithms, circumventing the need for a very long quantum circuit for the quantum state exponentiation \cite{bromley2019batched}, which has been deemed intractable even for far-term quantum computers when compiled \cite{scherer2017concrete}. This does not remove the requirement of the state preparation (which requires QRAM or a special oracle \cite{giovannetti2008quantum}). Though, in the case of quantum data, this method does not need access to such exotic components, and has the potential to demonstrate a quantum advantage for learning the unitary which diagonalizes either a quantum Hamiltonian or quantum density matrix.

Future iterations on this work could modify the loss function, potentially introducing a quantum variant of the Evidence Lower Bound (ELBO) for the quantum relative entropy loss. This would, in turn, yield a quantum form of Variational Autoencoders (VAEs) \cite{kingma2013auto}. Both QHBMs and VAEs could potentially be used to learn an effective mixed state representation of an unknown quantum state from partial quantum tomographic data~\cite{d2001quantum}. We aim to tackle this Quantum Neural State Tomography with QHBMs in future work.

In a near-term future iteration, we plan to fully combine classical neural network-based Energy-Based Models \cite{du2019implicit} with our QHBM. Recent papers have demonstrated the power of hybridized quantum-classical neural networks trained using hybrid quantum-classical backpropagation, originally introduced in \cite{verdon2018universal}, then later iterated upon in \cite{schuld2019evaluating}. Given previous successes in combining classical differentiable models and quantum variational circuits, we anticipate this implementation to be straightforward given the framework laid out in this paper. Since such a full hybridization would allow us to go beyond full-rank multinoulli estimation for the general classical latent distribution QHBM, this would be the piece which would unlock the scalability of this algorithm to highly non-trivial large-scale quantum systems. We plan to explore this in upcoming work.

QHBM present several key advantages relative to other forms of unsupervised quantum learning with quantum neural networks such as quantum GANs~\cite{Lloyd2018qgan}. GANs (both quantum and classical) are notoriously difficult to train, and once trained, difficult to extract physical quantities from. QHBM represent physical quantities more directly, and as such are much more suitable for applications that involve physical quantum data. From our preliminary numerics, our framework also appears to train very robustly and with few iterations. Furthermore, QHBM require less quantum circuit depth during training than quantum GANs, which require both a quantum generator and quantum discriminator. The latter is a key consideration for possible implementation on near-term intermediate scale quantum devices \cite{preskill2018quantum}.

A key point is that in Quantum Modular Hamiltonian Learning (QMHL) and VQT, we are not just learning a distribution for the modular Hamiltonian and thermal state respectively. Rather, we are learning efficient approximate parameterizations of these quantities with our quantum circuit. In the context of VQT, this gives us the ability to directly prepare a thermal state on our quantum computer using polynomial resources, from knowledge of a corresponding Hamiltonian. For Modular Hamiltonian Learning, once an efficient ansatz has learned the optimal value of parameters such that its output approximates our data mixed state, this gives us the ability to reproduce as many copies of the learned mixed state distributions as desired. In addition, the modular Hamiltonian itself provides invaluable information related to topological properties, thermalization, and non-equilibrium dynamics.

In particular, Modular Hamiltonian Learning gives one access to the eigenvalues of the density matrix and the unitary that diagonalizes the Modular Hamiltonian. Applying the QNN to a quantum state brings it into the eigenbasis of the Modular Hamiltonian. In this basis, an exponentiation of the diagonal latent modular Hamiltonian (which we have a classical description of) implements modular time evolution. We can apply the inverse QNN to return to the original computational basis.

In the same vein, QMHL provides the ability to probe a system at different temperatures, something that mixed state learning on its own does not. Given access to samples from a thermal state at some temperature, one can thus generate typical samples from the same system at another temperature by learning the modular Hamiltonian and systematically changing the latent space parameters.

Finally, we underscore the generality of our framework. Just as classical EBM are inspired by physics but can be applied to problems far abstracted from real physical systems, so too can our QHBM be employed for any task which involves learning a quantum distribution.

\section{Acknowledgements}

Numerical experiments in this paper were executed using a custom combination of Cirq \cite{cirq}, TensorFlow \cite{tensorflow2015-whitepaper}, and OpenFermion \cite{openfermion}. GV, JM, and SN would like to thank the team at X for the hospitality and support during their respective Quantum@X residencies where this work was completed. X, formerly known as Google[x], is part of the Alphabet family of companies, which includes Google, Verily, Waymo, and others (\url{www.x.company}). GV acknowledges funding from NSERC. 


\bibliography{references}

\begin{appendix}

\section{Quantum Neural Networks and Gradients}
\label{app:qnn}

A Quantum Neural Network can generally be written as a product of layers of unitaries in the form
\begin{align}\label{eq:full_par_circ}
    \hat{U}(\bm{\phi}) = \prod_{\ell=1}^L\hat{V}^{\ell}\hat{U}^{\ell}(\bm{\phi}^{\ell}),
\end{align}
where the $\ell^\text{th}$ layer of the QNN consists of the product of $\hat{V}^{\ell}$, a non-parametric unitary,  and $\hat{U}^{\ell}(\bm{\phi}^{\ell})$ a unitary with (possibly) multiple variational parameters (note the superscripts here represent indices rather than exponents). The multi-parametric unitary of a given layer can itself be generally comprised of multiple unitaries $\{\hat{U}_{j}^{\ell}(\phi^{\ell}_{j})\}_{j=1}^{M_\ell}$ applied in parallel:
\begin{align}
    \hat{U}^{\ell}(\bm{\phi}^{\ell})\equiv \bigotimes_{j=1}^{M_\ell} \hat{U}_{j}^{\ell}(\phi^{\ell}_{j}),
\end{align}
here $\mathcal{I}_{\ell}$ represents the set of indices of the corresponding to the $\ell^\text{th}$ layer.  Finally, each of these unitaries $\hat{U}_{j}^{\ell}$ can be expressed as the exponential of some generator $\hat{g}_{j\ell}$, which itself can be any Hermitian operator on $n$ qubits (thus expressible as a linear combination of $n$-qubit Pauli's),
\begin{align}\label{eq:Pauli_decomp}
  \hat{U}^{\ell}_{j}(\phi^{\ell}_{j})=   e^{-i\phi^{\ell}_{j} \hat{g}_j^{\ell }}, \quad \hat{g}_j^{\ell} = \sum_{k=1}^{K_{j\ell}} \beta^{j\ell}_k \hat{P}_k,
\end{align}
 where $\hat{P}_k \in \mathcal{P}_n$ (Paulis on $n$-qubits \cite{Gottesman1997}) and $\beta^{j\ell}_k\in \mathbb{R}$ for all $k,j,\ell$. For a given $j$ and $\ell$, in the case where all the Pauli terms commute, i.e.  $[\hat{P}_k,\hat{P}_m]=0$ for all $m,k$ such that $\beta^{j\ell}_m,\beta^{j\ell}_k\neq 0$, one can simply decompose the unitary into a product of exponentials of each term,
 \begin{align}\label{eq:}
    \hat{U}^{\ell}_{j}(\phi^{\ell}_{j}) = \prod_k e^{-i\phi^{\ell}_{j} \beta^{j\ell}_k \hat{P}_k}.
 \end{align}
Otherwise, in instances where the various terms do not commute, one may apply a Trotter-Suzuki decomposition of this exponential \cite{suzuki1990fractal}, or other quantum simulation methods \cite{campbell2019random}.

 In order to optimize the parameters of an ansatz from equation \eqref{eq:full_par_circ}, we need a cost function to optimize. In the case of standard variational quantum algorithms this cost function is most often chosen to be the expectation value of a cost Hamiltonian, 
 \begin{equation}\label{eq:cost}
     f(\bm{\phi}) = \braket{\hat{H}}_{\bm{\phi}} \equiv  \bra{\Psi_0}\hat{U}^\dagger(\bm{\phi})\hat{H}\hat{U}(\bm{\phi})\ket{\Psi_0}
 \end{equation}
where $\ket{\Psi_0}$ is the input state to the parametric circuit. In general, the cost Hamiltonian can be expressed as a linear combination of operators, e.g. in the form
 \begin{equation}\label{eq:cost_lincomb}
     \hat{H} = \sum_{k=1}^N \alpha_k\hat{h}_k \equiv \bm{\alpha}\cdot \bm{\hat{h}},
 \end{equation}
 where we defined a vector of coefficients $\bm{\alpha}\in \mathbb{R}^N$ and a vector of $N$ operators $\bm{\hat{h}}$.
 Often this decomposition is chosen such that each of these sub-Hamiltonians are in the $n$-qubit Pauli group $\hat{h}_k \in\mathcal{P}_n$. The expectation value of this Hamiltonian is then generally evaluated via quantum expectation estimation, i.e. by taking the linear combination of expectation values of each term 
\begin{equation}\label{eq:qee_lincomb}
     f(\bm{\phi}) =\braket{\hat{H}}_{\bm{\phi}} = \sum_{k=1}^N \alpha_k \braket{\hat{h}_k}_{\bm{\phi}} \equiv \bm{\alpha}\cdot \bm{h}_{\bm{\phi}},
 \end{equation}
 here we introduced the vector of expectations $\bm{h}_{\bm{\phi}} \equiv \braket{\bm{\hat{h}}}_{\bm{\phi}} $.
 In the case of non-commuting terms, the various expectation values $\braket{\hat{h}_k}_{\bm{\phi}}$ terms are estimated over separate runs. 
 
 Now that we have established how to evaluate the loss function, let us describe how to obtain gradients of the cost function with respect to the parameters. A simple approach is to use simple finite-difference methods, for example, the central difference method, 
 \begin{equation}\label{eq:central_diff}
     \partial_k f(\bm{\phi}) = \tfrac{1}{2\varepsilon}[f(\bm{\phi} + \varepsilon\bm{\Delta}_k)-f(\bm{\phi} - \varepsilon\bm{\Delta}_k)] +\mathcal{O}(\varepsilon^2)
 \end{equation}
 which, in the case where there are $M$ continuous parameters, involves $2M$ evaluations of the objective function, each evaluation varying the parameters by $\epsilon$ in some direction, thereby giving us an estimate of the gradient of the function with a precision $\mathcal{O}(\varepsilon^2)$. Here the $\bm{\Delta}_k$ is a unit-norm perturbation vector in the $k^{\text{th}}$ direction of parameter space, $(\bm{\Delta}_k)_j = \delta_{jk}$. In general, one may use lower-order methods, such as forward difference with $\mathcal{O}(\varepsilon)$ error from $M+1$ objective queries \cite{farhi2018classification}, or higher order methods, such as a five-point stencil method, with $\mathcal{O}(\epsilon^4)$ error from $4M$ queries \cite{abramowitz20061965}. 
 
 As recently pointed out in various works \cite{schuld2018evaluating,harrow2019low}, given knowledge of the form of the ansatz (e.g. as in \eqref{eq:Pauli_decomp}), one can measure the analytic gradients of the expectation value of the circuit for Hamiltonians which have a single-term in their Pauli decomposition \eqref{eq:Pauli_decomp} (or, alternatively, if the Hamiltonian has a spectrum  $\{\pm \lambda\}$ for some positive $\lambda$). For multi-term Hamiltonians, in \cite{schuld2018evaluating} a method to obtain the analytic gradients is proposed which uses a linear combination of unitaries. Recent methods use a stochastic estimation rule to reduce the number of runs needed and keep the optimization efficient \cite{harrow2019low}. In our numerics featured in this paper, we simply used first-order finite (central) difference methods for our gradients.

\section{Gradients of QHBMs for generative modelling}
\label{app:qhbm_grad}

\subsection{Gradients of model parameters}

Using our notation for the \textit{pulled-back data state} 
\(\hat{\sigma}_{\mathcal{D},\bm{\phi}} \equiv \hat{U}^\dagger(\bm{\phi})\hat{\sigma}_{\mathcal{D}}\hat{U}(\bm{\phi})\),
the gradient with respect to model parameters $\bm{\phi}$ will be given by 
\begin{align}\partial_{\phi_j}\mathcal{L}(\bm{\theta,\phi}) = \partial_{\phi_j}\tr(\hat{K}_{\bm{\theta}}\hat{\sigma}_{\mathcal{D},\bm{\phi}})= \partial_{\phi_j}\braket{\hat{K}_{\bm{\theta}}}_{\hat{\sigma}_{\mathcal{D},\bm{\phi}}},\end{align}
simply is the gradient of the latent modular Hamiltonian expectation value with respect to the data state pushed through the reverse quantum neural network $U^\dagger(\bm{\phi})$. Taking gradients of the expectation value of a multi-term observable is the typical scenario encountered in regular VQE and quantum neural networks, there exist multiple strategies to get these gradients (see Appendix \ref{app:qnn}).

\subsection{Gradients of variational parameters}

For gradients of the loss function with respect to model parameters $\bm{\theta}$, note that we need to take gradients of both the modular Hamiltonian terms and the partition function. For gradients of the modular Hamiltonian expectation term with respect to the variational parameters $\bm{\theta}$, we change the observable we are taking the expectation value of, this is not as common in the literature on quantum neural networks, but it is fairly simple: 

\begin{align}\partial_{\theta_m}\braket{\hat{K}_{\bm{\theta}}}_{\hat{\sigma}_{\mathcal{D},\bm{\phi}}} =\braket{\partial_{\theta_m}\hat{K}_{\bm{\theta}}}_{\hat{\sigma}_{\mathcal{D},\bm{\phi}}} = \braket{\partial_{\theta_m}\hat{K}_m(\theta_m)}_{\hat{\sigma}_{\mathcal{D},\bm{\phi}}}, \end{align}
it is simply the expectation value of the gradient. Note that here we assumed the latent modular Hamiltonian was of the form
\begin{equation}
\hat{K}_{\bm{\theta}} = \sum_m \hat{K}_m (\bm{\theta}),
\end{equation}
which is valid both for a diagonal latent multinoulli variable ($\hat{K}_m = \ket{m}\!\bra{m}$) and a factorized latent distribution, in which case each $\hat{K}_m$ can be understood as a subsystem Hamiltonian.

For simple cases where the parameterized modular Hamiltonian is simply a scalar parameter times a fixed operator $\hat{M}$, i.e.,  $\hat{K}_m(\theta_m) = \theta_m \hat{M}_m$, then the above simplifies to 
\begin{align}\partial_{\theta_m}\braket{\hat{K}_{\bm{\theta}}}_{\hat{\sigma}_{\mathcal{D},\bm{\phi}}} = \braket{\hat{M}_m}_{\hat{\sigma}_{\mathcal{D},\bm{\phi}}} = \tr(\hat{M}_{m}\hat{\sigma}_{\mathcal{D},\bm{\phi}}),\end{align}
which is a simple expectation value.

Now, to get the gradient of the partition function, given our assumed knowledge of each of the simple modular Hamiltonian, we should have an analytical expression for the partition function which we can use. Notice,
\begin{align}\textstyle\log \mathcal{Z}_{\bm{\theta}}  &= \sum_j \log \mathcal{Z}_j(\theta_j)  \\& = \nonumber \sum_j\log[\tr(e^{-\hat{K}_j(\theta_j)})]  \\ &=\sum_j \nonumber \log(\sum_{k_j} k_j(\theta_j)) \end{align}
where $k_j(\theta_j)$ are the eigenvalues of the $j$th parameterized modular Hamiltonian $\hat{K}_j(\theta_j)$. 

We can use this form to derive the gradient of this term with respect to model parameters
\begin{align} \partial_{\theta_m}\log \mathcal{Z}_{\bm{\theta}} &= \partial_{\theta_m} \log[\tr(e^{-\hat{K}_m(\theta_m)})] \\&= \nonumber \tfrac{1}{\mathcal{Z}_m(\theta_m) } \partial_{\theta_m}\tr(e^{-\hat{K}_m(\theta_m)})\\&= \nonumber -\tfrac{1}{\mathcal{Z}_m(\theta_m) } \tr(e^{-\hat{K}_m(\theta_m)}\partial_{\theta_m}\hat{K}_m(\theta_m)).
\end{align}
Once again, for simple parameterized modular Hamiltonians of the form $\hat{K}_m(\theta_m) = \theta_m \hat{M}_m$, then this formula becomes
\begin{align}\partial_{\theta_m}\log \mathcal{Z}_{\bm{\theta}}=-\tfrac{1}{\mathcal{Z}_m(\theta_m) } \tr(e^{-\theta_m \hat{M}_m}\hat{M}_m) \end{align}
which is straightforward to evaluate analytically classically given knowledge of the spectrum of $\hat{M}_m$. Technically, there is no need to use the quantum computer to evaluate this part of the gradient.

Thus, we have seen how to take gradients of all terms of the loss function.
Thus, by simply training on the quantum cross entropy between the data and our model, we can theoretically fully train our generative model using gradient-based techniques.

\section{Distance Metrics}
\label{app:dist}

The trace distance is defined to be
\begin{align}
    T(\hat{\rho}, \hat{\sigma}) := \frac{1}{2} Tr[\sqrt{(\hat{\rho} - \hat{\sigma})^{\dag}(\hat{\rho} - \hat{\sigma})}],
\end{align}

and generic mixed state fidelity is calculated as 

\begin{align}
F(\hat{\rho}, \hat{\sigma}) := \Big[ Tr \sqrt{\sqrt{\hat{\rho}} \hat{\sigma} \sqrt{\hat{\rho}}} \Big]^2
\end{align}

\section{Investigation of Ansatz Performance}
\label{app:ansatz}
In the case of Gaussian states, we are guaranteed that a factorized latent space solution exists. For qubit systems with inter-qubit coupling, this is not the case. Of course fully parameterizing both the unitary and the classical distribution should allow for perfect reconstruction - and hybrid quantum-classical algorithms should enable the efficient extraction of typical samples. That said, it is interesting to see just how much representational power the product ansatz actually has when applied to non-Gaussian systems. To this end, we investigate its performance both as a function of temperature, and in comparison to the actual performance of the general ansatz, which has a fully parameterizable diagonal distribution.

\begin{figure}[ht]
\centering
\includegraphics[width=0.45\textwidth]{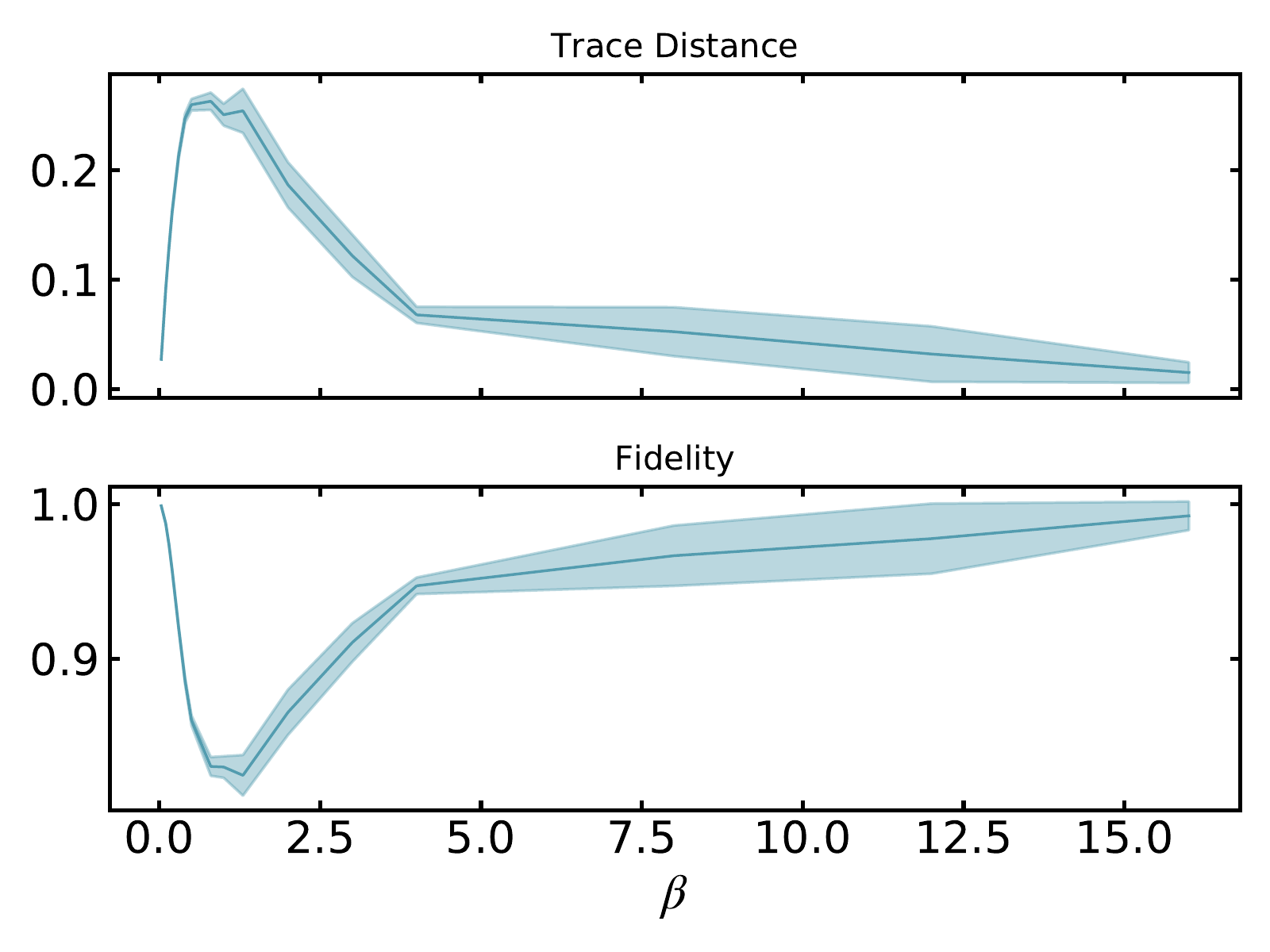}
\caption{Performance metrics for QMHL with factorized latent space ansatz on the thermal state of a fixed Hamiltonian at different temperatures. Solid dark line denotes mean, and shaded translucent region denotes $95\%$ confidence interval. Model under consideration is a four qubit one-dimensional Heisenberg model $J = -1.0$ with transverse and longitudinal magnetic fields, $J_x = 0.3, \quad J_z = 0.2$, and $50$ trial initializations were chosen for each inverse temperature $\beta$.}
\label{fig:betadep}
\end{figure}

Our simulations are small in scale and restricted in the classes of models considered, but promising nonetheless. In Figure~\ref{fig:betadep}, we show trace distance and fidelity for a one-dimensional heisenberg model with transverse and longitudinal magnetic fields,

\begin{align}
\hat{H} = -\sum_{\langle ij\rangle} \hat{S}_i\cdot \hat{S}_j + \sum_j (h^x S^x_j + h^z S^z_j),
\end{align}

where the presence of perpendicular field terms induces quantum fluctuations. In the limit $\beta \rightarrow 0$, the thermal fluctuations completely dominate the quantum, and the resulting thermal state is completely mixed. The essentially perfect reconstruction in that limit - and with little variance - reflects the high approximate degeneracy of the solution space. In the opposite limit, $\beta \rightarrow \infty$, the thermal state approaches the ground state. Classical correlations vanish and the state becomes pure. In this limit, we achieve asymptotically ideal reconstruction with high probability, but occasionally the variational procedure gets stuck in local minima, resulting in the high variance around the mean. 

In the intermediate regime both quantum and classical correlations are important, and the lack of generality in factorized latent space results in slight performance losses with respect to a generic ansatz. For large systems, the fidelity and trace distance will not necessary perform well on models trained to optimize relative entropy.

\begin{table}
\centering
\caption{Reconstruction metrics for QMHL with latent space product (factorized) and general ansatz, computed for a family of random coupling models on a four qubit chain after convergence or 200 training steps. For each of the 10 model Hamiltonians, 10 random initializations were chosen for each ansatz. Mean, min and max are with respect to these initializations, and the values reported represent averages over the class of Hamiltonians.}
\subtable[Fidelity]{
\label{tab:fidelity}
\begin{tabular}{llll}
\hline
\textbf{Ansatz} & \textbf{Mean} & \textbf{Min} & \textbf{Max} \\ \hline
Factorized & 0.871 & 0.752 & 0.919 \\ \hline
General & 0.935 & 0.881 & 0.950 \\ \hline
\end{tabular}
}

\subtable[Trace Distance]{
\label{tab:tracedist}
\begin{tabular}{llll}
\hline
\textbf{Ansatz} & \textbf{Mean} & \textbf{Min} & \textbf{Max} \\ \hline
Factorized & 0.221 & 0.172 & 0.367 \\ \hline
General & 0.173 & 0.153 & 0.249 \\ \hline
\end{tabular}
}
\end{table}

In the reconstruction tables for fidelity~\ref{tab:fidelity} and trace distance,~\ref{tab:tracedist}, QMHL was performed at inverse  temperature $\beta = 1.3$, which is in the general region where we expect most drastic deviation from the factorized latent space assumption. We see that the performance loss as compared to the generic ansatz is relatively small. Random models were chosen from the class
\begin{align}
\hat{H} = \sum_{\langle ij\rangle} (\alpha_{ij} \hat{X}_i \hat{X}_j + \beta_{ij} \hat{Y}_i \hat{Y}_j + \gamma_{ij} \hat{Z}_i \hat{Z}_j) + \sum_j \bm{h}_j \cdot \hat{\bm{S}}_j,
\end{align}
where $\alpha_{ij}, \beta_{ij}, \gamma_{ij}, h_j$ were all sampled independently from the uniform distribution on the interval $[-1, 1]\subset{\mathbb{R}}$, and allowed to differ on each site and edge.

\section{Continuous Variable Gaussian Quantum Information}
\label{app:cv}

By Wick's theorem, Gaussian states (Bosonic or Fermionic) are completely specified by their first and second moments. Mathematically, this means that all higher order correlation functions ($n$-point correlators) can be written as sums of products of second order correlators:

\begin{align}
    \langle \hat{\xi}^{a_1} \cdots \hat{\xi}^{a_n}\rangle = \sum \prod  \langle \hat{\xi}^{a_j} \cdots \hat{\xi}^{b_j}\rangle
\end{align}

For thermal states, the mean displacement is zero and the density matrix $\hat{\rho}$ is completely replaced by the covariance matrix $\Gamma$ ($2 N\times 2N$ for systems of $N$ Bosons).

The `position' and `momentum' quadratures, defined by 

\begin{align}
    \hat{x}_{j} & = \frac{1}{\sqrt{2}} (\hat{a}_j + \hat{a}^{\dag}_j) \\
    \hat{p}_{j} & =  \frac{\mathrm{i}}{\sqrt{2}} (\hat{a}_j - \hat{a}^{\dag}_j),
\end{align}

are conjugate variables, satisfying the canonical Bosonic commutation relations, $ [\hat{x}_j, \hat{p}_k] = 2\mathrm{i} \delta_{jk}$. If we combine all of the quadratures into a vector of operators, $\bm{\xi} = (\hat{x}_1, \ldots, \hat{x}_N, \hat{p}_1, \ldots, \hat{p}_N)$, then the Bosonic commutations take the form

\begin{align}
    [\xi^a, \xi^b] = \mathrm{i} \Omega_{ab},
\end{align}

where \begin{align}\Omega = \begin{bmatrix}
 0 & I_N \\
-I_N & 0
\end{bmatrix}\end{align}
is a $2N \times 2N$ skew-symmetric matrix known as the symplectic form, and $I_N$ is the $N \times N$ identity matrix.

Bosonic Gaussian transformations are then transformations that preserve the symplectic form, i.e. $S \in Sp (2 N_b)$ where

\begin{align}
    S^T \Omega S = \Omega.
\end{align}

The uncertainty principle (encoded in the commutation relations) constrains the covariance matrix to $\Gamma_{B} \geq i \Omega$.

In symplectic diagonal form, we now have independent effective oscillators, and the entropy takes the simplified form  

\begin{align}
S_{B}(\sigma) = \sum_j \frac{\omega_j}{e^{\omega_j} - 1 } - \mathrm{log} (1 - e^{- \omega_j}),
\end{align}

where $\omega_j$ is the frequency of the $j^{\mathrm{th}}$ effective oscillator, and $\beta$, $\hbar$, and $k_B$ are set to unity.

The logarithm of the partition function also takes a simplified form:

\begin{align}
    \mathrm{log}\mathcal{Z}_{\bm{\theta}} = \sum_j - \frac{\omega_j}{2} - \mathrm{log}(1 - e^{- \omega_j}),
\end{align}

\section{Gaussian Model Parameterizations}
\label{app:param}
In Section~\ref{sec:apps_n_exps}, we utilize the Gaussian nature of the problems to better parameterize the unitary transformations. Here, we explain those parameterizations and how one may operationally implement such parameterized transformations on various quantum computing platforms.

\subsection{Fermions}
As mentioned in the main text, Fermionic Gaussian transformations are given by orthogonal transformations acting on Fermionic modes. Multiple methods exist for parameterizing the class of orthogonal transformations. In classical simulation of QHBM, one can take advantage of a consequence of the lie algebra - lie group correspondence: the space of real orthogonal matrices of size $N$ is spanned by $e^{A}$ paremeterized by anti-symmetrizing a lower-diagonal matrix with $N (N - 1)/2$ non-zero entries. 

For QHBM on a real quantum computer, it is preferable to use Givens rotations on pairs of modes. The matrix describing this two-mode mixing is a rotation, 

\begin{align}
    G(\phi) = \begin{pmatrix}
 \cos(\phi) & -\sin(\phi) \\
 \sin(\phi) & \cos(\phi)
\end{pmatrix}.
\end{align}

Such a rotation can be generalized to include complex phases, but that is unnecessary when staying within the confines of the special orthogonal group. The space of such orthogonal transformations can be systematically decomposed into sequences of Givens rotations which act on pairs of modes, as detailed in ~\cite{PhysRevApplied.9.044036}.

\subsection{Bosons}
By the Bloch-Messiah decomposition~\cite{gosson2011}, any symplectic matrix $S$ of size $2N$ can be brought into the form 
\begin{align}
    S = O_1 D O_2
\end{align}

where $O_1$ and $O_2$ are real orthogonal and symplectic, and D is a diagonal matrix of the form $\mathrm{diag}(d_1, d_2, \ldots d_N, d_1^{-1}, d_2^{-1}, ..., d_N^{-1})$.

For classical simulation of QHBM, it is useful to decompose this further. Every real orthogonal symplectic matrix, 
\begin{align*}
O = \begin{bmatrix}
 X & Y \\
-Y & X
\end{bmatrix} \in \mathcal{R}^{2N \times 2N}
\end{align*}

can be associated to a unitary $U = X + i Y \in \mathcal{C}^{N\times N}$. The space of such unitaries is spanned by exponentiating anti-Hermitian matrices. Each Hermitian matrix is described by $N ( N - 1)/2$ complex parameters. Finally, we can generate Hermitian matrices by anti-symmetrizing (with hermitian conjugate) complex-valued lower-diagonal matrices. We can thus classically parameterize the space of symplectic transformations by reading the sequence of decompositions backwards starting from lower-diagonal matrices. Altogether, a symplectic matrix for an $N$-mode system may be described by $2 N^2 - N$ real parameters. 

In quantum optical setups, $O_1$ and $O_2$ represent passive transformations, such as performed by beam splitters and phase shifters, and $D$ can be implemented via a sequence of single-mode squeezers, as explained in ~\cite{qcvbook}.

\section{Gaussian Fermionic Quantum Systems}
\label{app:gaussfermion}

In this appendix, we give a more detailed prescription for how one could implement QHBM for Fermionic Gaussian systems. While we employ the Jordan-Wigner transformation~\cite{jw1, jw2} to symbolically map Fermionic operators to qubit operators, we note that other mappings, like Bravyi-Kitaev (BK)~\cite{bk1, bk2} or Ball-Verstraete-Cirac (BVC)~\cite{bvc1, bvc2, bvc3} transformations also suffice as long as appropriate parameterizations are chosen.

In the JW transformation, effective Fermionic creation and annihilation operators are induced by combining qubit operators, 

\begin{align}
\hat{a}_k & = \frac{1}{2} (\hat{X}_k + i \hat{Y}_k) \hat{Z}_1 \cdots \hat{Z}_{k-1} \nonumber \\
\hat{a}^{\dag}_k & = \frac{1}{2} (\hat{X}_k - i \hat{Y}_k) \hat{Z}_1 \cdots \hat{Z}_{k-1},
\end{align}

where $\hat{X}_k$, $\hat{Y}_k$, and $\hat{Z}_k$ are Pauli operators on qubit $k$. In this language, it's natural to consider the occupation of the $k^{\mathrm{th}}$ effective Fermionic mode, 

\begin{align}
\hat{n}_k \quad \equiv \quad \hat{c}^{\dag}_k \hat{c}_k \quad = \quad \frac{1}{2} (1 - \hat{Z}_k),
\end{align}

and we see that we can 'measure' the occupation of the mode just by measuring the Pauli-Z operator on the corresponding qubit.

In making this prescription, we are implicitly defining an ordering on the qubits. For one-dimensional systems, this is natural, and all next neighbor effective Fermionic hopping gates are local in qubit operators:

\begin{align}
\hat{c}^{\dag}_k \hat{c}_{k + 1} + \hat{c}^{\dag}_{k+1} \hat{c}_{k} = \frac{1}{2} (\hat{X}_k \hat{X}_{k + 1} + \hat{Y}_k \hat{Y}_{k + 1}).
\end{align}

For higher-dimensional systems, methods have been developed that handle the extra phases involved.

For each pair of physical Fermionic creation and annihilation operators, $\hat{a}_j, \hat{a}^{\dag}_j$, we can associate a pair of virtual Majorana modes, $\hat{c}_{2j - 1}, \hat{c}_{2j}$, according to

\begin{align}
    \hat{c}_{2j-1} & = \hat{a}_j + \hat{a}^{\dag}_j \\
    \hat{c}_{2j} & = \mathrm{i} (\hat{a}_j - \hat{a}^{\dag}_j)
\end{align}

The canonical transformation from physical Fermionic operators to Majorana operators is a local rotation in operator space for each Fermion individually. While the defining equations for the Majoranas are cosmetically identical (up to a multiplicative constant) to those defining Bosonic conjugate variables $\hat{p}$ and $\hat{q}$, these Majorana Fermions satisfy the canonical Fermionic anti-commutation relations, $\{ \hat{c}_k, \hat{c}_l\} = 2\delta_{kl}$.

In the Majorana basis, any Fermionic Gaussian Hamiltonian takes the generic form 
\begin{align}
    \hat{H}_{F} = i\sum_{i, j}^{2N_f}h_{ij} \hat{c}_i\hat{c}_j + E,
\end{align}

where $N_f$ is the total number of Fermions, and $E$ is some constant energy shift associated with the transformation to Majorana modes.

In this basis, we can approach the Fermionic problem in a similar manner to the Bosonic problem - we want to decompose the system (now written in terms of Majoranas) into independent effective Fermionic oscillators. Said in another way, we want to find a set of Fermionic operators, $\{\hat \tilde{c}_1, \ldots, \tilde{c}_{2N_f}  \}$ and a corresponding set of energies $\{\epsilon_1, \ldots, \epsilon_{2N_f} \}$, such that 

\begin{align}
\hat{H}_{F} = \sum_{j = 1}^{2 N_f} \epsilon_j  \tilde{c}^{\dag}_j \tilde{c}_j + \mathrm{const.}
\end{align}

As mentioned in the main text, this diagonalization can be performed by an orthogonal transformation on the covariance matrix associated with the thermal state. 

In general, an orthogonal transformation on Fermionic modes must also contain particle-hole transformations, but working in the Majorana basis these are not necessary. A generic orthogonal matrix requires no more than $\mathcal{O}(N_f^2)$ such Givens rotations.

In one-dimensional systems, the JW transformation makes for a remarkably simple implementation of Givens rotations, including two CNOT gates and one controlled-phase rotation. For higher-dimensional systems, techniques exist for dealing with the phases that arise in the JW mapping. Moreover, Givens rotations are incredibly close to the physical operations performed on real quantum devices. 

Finally, the derivative of a Givens rotation analytically is 

\begin{align}
    G'(\phi) = G( \phi + \frac{\pi}{2}),
\end{align}

meaning that gradients can be computed easily using exact methods, by physically measuring expectation values with parameters shifted by $\frac{\pi}{2}$.

\section{D-wave Superconductivity}
\label{app:dwave}

The mean-field BCS Hamiltonian has become a paradigmatic phenomenological model in the study of superconductors.

The generic model is described by Hamiltonian

\begin{align}
  \hat{H}_{d_{x^2-y^2}} = & -t \sum_{\langle i,j\rangle, \sigma}( \hat{a}^{\dag}_{i, \sigma} \hat{a}_{j, \sigma}  + \hat{a}^{\dag}_{j, \sigma} \hat{a}_{i, \sigma})  + \mu \sum_{i, \sigma}\hat{a}^{\dag}_{i, \sigma} \hat{a}_{i, \sigma}  \nonumber \\
  & \sum_{\langle i,j\rangle} \Delta_{ij} (\hat{a}^{\dag}_{i, \uparrow} \hat{a}^{\dag}_{j, \downarrow} - \hat{a}^{\dag}_{i, \downarrow} \hat{a}^{\dag}_{j, \uparrow} + \hat{a}_{j, \downarrow} \hat{a}_{i, \uparrow} - \hat{a}_{j, \uparrow} \hat{a}_{i, \downarrow})
\end{align}

defined for Fermions (electrons) on a two-dimensional lattice of size $N_x$ by $N_y$ sites, where $\hat{a} \, (\hat{a}_j^{\dag})$ is a Fermionic annihilation (creation) operator on lattice site $j$. Each site hosts two spin-orbitals (one for spin up, and one for spin down), giving a total of $2 \times N_x \times N_y$ Fermions. Here, $\mu$ represents a chemical potential that sets the filling density. $t$ denotes next-neighbor hopping strength, and $\Delta_{ij}$ parameterizes the superconducting gap (less formally, it incentivizes the formation of Cooper pairs).

\end{appendix}
\end{document}